\documentclass[11pt]{article}
\usepackage{amsfonts}
\usepackage{amsmath}
\usepackage{amssymb}
\usepackage{bbm}
\usepackage{threeparttable}

\usepackage{natbib}

\usepackage{graphicx}
\usepackage{lscape}
\usepackage[justification=centering]{caption}
\usepackage{dcolumn}
\usepackage{booktabs}
\usepackage{tabularx}
\usepackage{colortbl}
\usepackage{arydshln}
\usepackage{comment}
\usepackage{todonotes}
\usepackage[hidelinks]{hyperref}
\usepackage{afterpage}
\usepackage{dcolumn}
\usepackage{scrextend}
\usepackage{subcaption}
\usepackage{xcolor}
\usepackage{centernot}
\usepackage{pgf}
\usepackage{setspace}
\usepackage{url}
\usepackage{pdfprivacy}

\presetkeys{todonotes}{size=\footnotesize}{}
\renewcommand{\baselinestretch}{1.38}
\newcommand{\lspace}[1]{\renewcommand{\baselinestretch}{#1} \small\normalsize}

\newcommand{\eu}{\operatorname{e-u}}

\newcommand{\att}[1]{\tau^{#1}}
\newcommand{\attest}[1]{\widehat{\tau}^{#1}}

\def\ci{\mathrel{\perp\mspace{-10mu}\perp}}

\def\E{\mathbb{E}}

\begin{document}
\lspace{1}

\vspace{-.5in}

\title{Close Enough? A Large-Scale Exploration of Non-Experimental Approaches to Advertising Measurement\thanks{\enspace Data was de-identified such that consumers could not be identified and was analyzed in aggregate when possible. We thank Neha Bhargava, JP Dub\'e, Dean Eckles, Paul Ellickson, Garrett Johnson, Randall Lewis, Oded Netzer, Kelly Paulson, Julian Runge, and seminar participants at Facebook, the Marketing Science Conference, the Conference on Digital Experimentation, the NYU-Temple-CMU Conference, and the Virtual Quantitative Marketing Seminar. E-mail addresses for correspondence: b-gordon@kellogg.northwestern.edu, f-zettelmeyer@kellogg.northwestern.edu, rmoakler@meta.com.}}

\author{Brett R.~Gordon \\
Kellogg School of Management\\
Northwestern University\\
\medskip \and Robert Moakler \\
Ads Research\\
Meta \medskip \and Florian Zettelmeyer \\
Kellogg School of Management\\
Northwestern University and NBER\\
}

\date{September 21, 2022}

\maketitle
\thispagestyle{empty}
\lspace{1.1}

\begin{abstract}

% Original
%\noindent Randomized controlled trials (RCTs) have become increasingly popular in both marketing practice and academia. However, RCTs are not always available as a solution for advertising measurement, necessitating the use of observational methods. We present the first large-scale exploration of two observational methods, double/debiased machine learning (DML) and stratified propensity score matching (SPSM). Specifically, we analyze 663 large-scale experiments at Facebook, each of which is described using over 5,000 user- and experiment-level features. Although DML performs better than SPSM, neither method performs well, despite using deep learning models to implement the propensity scores and outcome models. The median RCT lifts are 29\%, 18\%, and 5\% for the upper, middle, and lower funnel outcomes, respectively. Using DML (SPSM), the median lift by funnel is 83\% (173\%), 58\% (176\%), and 24\% (64\%), respectively, indicating significant relative measurement errors. We further leverage our large sample of experiments to characterize the circumstances under which each method performs comparatively better. However, broadly speaking, our results suggest that the structure and type of non-experimental campaign and user-level data logged at Facebook are inadequate to enable state-of-the-art observational approaches to reliably estimate an ad campaign's causal effect.

% New
\noindent Despite their popularity, randomized controlled trials (RCTs) are not always available for the purposes of advertising measurement. Non-experimental data is thus required. However, Facebook and other ad platforms use complex and evolving processes to select ads for users. Therefore, successful non-experimental approaches need to ``undo’’ this selection. We analyze 663 large-scale experiments at Facebook to investigate whether this is possible with the data typically logged at large ad platforms. With access to over 5,000 user-level features, these data are richer than what most advertisers or their measurement partners can access. We investigate how accurately two non-experimental methods---double/debiased machine learning (DML) and stratified propensity score matching (SPSM)---can recover the experimental effects. Although DML performs better than SPSM, neither method performs well, even using flexible deep learning models to implement the propensity and outcome models. The median RCT lifts are 29\%, 18\%, and 5\% for the upper, middle, and lower funnel outcomes, respectively. Using DML (SPSM), the median lift by funnel is 83\% (173\%), 58\% (176\%), and 24\% (64\%), respectively, indicating significant relative measurement errors. We further characterize the circumstances under which each method performs comparatively better. Overall, despite having access to large-scale experiments and rich user-level data, we are unable to reliably estimate an ad campaign's causal effect.

\vspace{.2in}
\noindent

\noindent  \textbf{Keywords}: Digital Advertising, Field Experiments, Causal Inference, Observational Methods, Advertising Measurement, Double ML.

\end{abstract}

\newpage
\setcounter{page}{1}
\lspace{1.3}
\section{Introduction}

In recent years, randomized controlled trials, or RCTs, have become increasingly popular in marketing practice and academia. This trend follows three important developments. First, many firms have invested heavily in their experimentation (and other analytical) capabilities, recognizing that RCTs are the gold standard of measurement \citep{kohavi_book}. Second, several leading advertising platforms have created experimentation tools that enable RCTs at no cost to advertisers.\footnote{Google: \url{https://www.thinkwithgoogle.com/intl/en-gb/marketing-resources/data-measurement/a-revolution-in-measuring-ad-effectiveness/}. Facebook: \url{https://www.facebook.com/business/help/552097218528551}. Microsoft: \url{https://help.ads.microsoft.com/\#apex/3/en/56908/-1}.} Finally, marketing academics have increasingly used RCTs to execute their research agendas \citep{LewisRaoReiley2015b}.

For advertising measurement, however, RCTs are not always available as a solution. Advertisers may operate under internal pressure to forgo a control group to maximize a campaign's reach. RCTs can also be technically difficult or even impossible to implement on many ad platforms \citep{johnson2022}. These are among the reasons why data scientists at advertisers, their third-party measurement partners, and advertising platforms have looked to alternative solutions for causal inference. The starting point for such an analysis is when an advertiser---instead of running an RCT---follows the more common practice of running a regular ad campaign. After choosing a population to target, typically only a subsample of these eligible users are eventually exposed to the ad campaign. To estimate the treatment effect, data scientists extract various measures from the exposed group (and potentially the unexposed group) and use these to estimate the causal or ``incremental'' effect of the ad campaign.

Facebook and other ad platforms use complex ranking and delivery processes to determine which ad, among all ads for which advertisers have placed bids, will be shown. The exact nature of the ad delivery process is unknown to bidders but usually takes into account the bid amount, the estimated click-through or conversion rate, and a relevance penalty to ensure that users are only shown ads that the advertising platform deems relevant to them. These features are continuously updated over time for each user and may be different for each new auction. As a result, selection into advertising exposure is based on a complex process powered by high-dimensional data. Successfully estimating the causal effect on an ad campaign using non-experimental data therefore requires that we undo the selection induced by this delivery process. This problem is difficult because the selection process uses auction-level data that advertising platforms typically do not log for future ad measurement analysis.

In this paper we investigate whether we can come ``close enough'' using the typical data stored on a large ad platform. To answer this question we analyze 663 ad experiments run between November 2019 and March 2020 on Facebook. The data contain approximately 7.9 billion user-experiment observations and over 38 billion ad impressions. These experiments were chosen to be representative of the large-scale experiments advertisers run on Facebook in the United States. The median ad experiment ran for 30 days with about 7.3 million users across the test and control groups. Within the test group, 77\% of users were exposed to at least one ad impression, and the median campaign accrued over 22 million impressions. These experiments represent a range of industry verticals such as E-commerce, Retail, Travel and Entertainment/Media. Many of these experiments measure several different conversion outcomes across a ``purchase funnel,'' such as page views (upper funnel), adding an item to a digital shopping cart (mid funnel), and purchase (lower funnel).

We estimate the causal effect of advertising using double/debiased machine learning (DML) \citep{dml_2018}. Traditional machine learning techniques, while appealing for their flexibility and efficiency, can produce biased estimates of causal effects due to regularization and overfitting. DML corrects for the bias introduced by regularization through orthogonalization and removes the bias from overfitting through cross-validation. This technique has become popular for estimating causal effects in a variety of settings, including both in academia and industry.\footnote{As of this writing (8/2/2022), \cite{dml_2018} has 1,255 citations in Google Scholar. See \cite{athey_imbens_2019} for a general review on the relevance of machine learning methods for empirical research. In industry, the technique is used at Uber (\url{https://medium.com/teconomics-blog/using-ml-to-resolve-experiments-faster-bd8053ff602e}) and Microsoft (\url{https://medium.com/data-science-at-microsoft/causal-inference-part-2-of-3-selecting-algorithms-a966f8228a2d}), and see the industry case studies presented in the tutorial in \cite{kdd_tutorial_2021}.} We also evaluate the preferred program evaluation method from \cite{lift1}, stratified propensity score matching (SPSM). For both models, we employ highly flexible and scalable deep learning methods to estimate each model's underlying components. Adopting a scalable method is important given the size and number of experiments we study.

We estimate our models using an extensive set of campaign- and user-level data logged at Facebook. To obtain an unbiased estimate of the causal effect, DML and SPSM appeal to the unconfoundedness assumption. Loosely speaking, this assumption requires that a user's potential outcomes are independent of treatment status, conditional on the user's features. Having a rich set of relevant features is thus critical if this assumption is to hold. We use four groups of variables: (1) a dense set of descriptive user features (e.g., age, gender, number of friends, number of ad impressions in the last 28 days); (2) a sparse set of user interest features (e.g., cooking? movies? sports?); (3) estimated action rates (e.g., estimated probability of conversion given exposure); (4) prior campaign-related conversion activity (e.g., 30-day lagged outcomes). A number of these features vary over time within each user and likely reflect the intensity of their online browsing behavior, helping us to address the issue of activity bias \citep{LewisRaoReiley2011}. Other variables, such as the estimated action rates, are a major factor in determining the winner of ad auctions at Facebook.

Collectively, the experiment sample, methodology, and features help us answer \textit{whether} the non-experimental ad campaign data logged at Facebook is sufficient to ``undo'' the selection induced by the ranking and delivery process, and therefore to estimate the causal effect of advertising. Moreover, using a large and representative set of experiments allows us to characterize \textit{when} the data allows us to recover the causal effect of advertising. If there are characteristics of ad campaigns where the data perform well, data scientists may be able to utilize non-experimental approaches to measure the impact of their ad campaigns for some advertisers. 

We find that SPSM performs poorly, despite making use of an extensive set of user-level features and a sophisticated machine learning model to estimate the propensity score. DML is, on average, less upwardly biased than SPSM. However, the remaining bias is substantial. The median RCT lifts are 29\%, 18\%, and 5\% for the funnel outcomes, respectively. Using DML (SPSM), the median lift by funnel is 83\% (173\%), 58\% (176\%), and 24\% (64\%), respectively, indicating significant relative measurement errors.  

We find that, using the data logged at Facebook, the causal inference approaches we investigate  perform comparatively better for prospecting campaigns rather than for remarketing campaigns and when the counterfactual conversion rates are smaller. Prospecting campaigns tend to employ broader targeting rules, whereas remarketing campaigns restrict attention to narrower groups of users who probably already interacted with the advertiser. Both findings could be due to the fact that prospecting campaigns tend to have lower baseline conversion rates than remarketing campaigns, such that conversions in the test group are more likely incremental to the ad campaign. 

In addition, we observe improved performance of SPSM and DML when experiments have more users, when a smaller share of users in the test group are exposed, and when the propensity model performs better. These findings point towards the fact that an overall larger set of unexposed users provides a better ``candidate pool'' to help either model estimate counterfactual outcomes for the exposed users. However, even with a large set of users, the underlying predictive models do not achieve sufficient accuracy in differentiating between exposed and unexposed users. 

% BG -- original -- 7.29.22
%In summary, we find that the non-experimental campaign and user-level data regularly logged and stored at Facebook are inadequate to control for the selection induced by the advertising platform and therefore fails to enable observational approaches to reliably estimate an ad campaign's causal effect. We attribute the failure to the type and structure of the data because we are using observational approaches which have worked successfully in other domains. We conjecture that the data at the disposal of data scientists at advertisers, their third-party measurement partners, and advertising platforms would do no better due to the intricacies present in many online ad delivery systems: To the best of our knowledge, the granularity and detail of the data we use is close to the best available at the most sophisticated advertising platforms and exceeds what individual advertisers or their third-party measurement partners typically could access.

% BG -- new -- 7.29.22
In summary, we find that despite the granularity of the data available and the flexibility of the models employed, we are unable to adequately control for the selection effects induced by the advertising platform. This leads our non-experimental estimates of ad effects to be biased relative to those obtained from RCTs. We believe this is more of a ``data problem'' than a ``model problem.''  We conjecture that the data at the disposal of data scientists at advertisers, their third-party measurement partners, and advertising platforms would do no better due to the intricacies present in many online ad delivery systems: To the best of our knowledge, the granularity and detail of the data we use is close to the best available at the most sophisticated advertising platforms and exceeds what individual advertisers or their third-party measurement partners typically could access.

To build and scale observational methods for advertising measurement, ad platforms would likely need to fundamentally alter their data logging and retention practices. To see this, consider that the selection of users into exposed versus unexposed groups occurs when an auction is triggered to show an ad to a user. The non-experimental data would need to contain the features needed to model the probability that the ad of interest (1) participates in the auction, (2) wins the auction, conditional on participating in the auction, and (3) is actually impressed, conditional on winning the auction. Moreover, since an auction takes place each time there's an opportunity to show an ad to someone, a platform would need to log features at the auction level and retain them in such a way to enable looking back over completed campaigns to obtain estimates of a campaign's effect. 

Advertising platforms typically do not log all of these data to enable future ad measurement. To understand why, consider two reasons we believe a company may not: First, processing the data required to deliver ads at a large scale involves many distributed servers, systems and engineering teams. Building out the required infrastructure to log these data as well as the state of each user's descriptive features for each ad auction, to enable campaign-level measurement, would be highly complex and require large investments. Second, companies may limit the storage of detailed data associated with individual-level impressions, and so storing data at the impression-bid level would be even more difficult. Using Facebook as an example, given the nearly two billion daily active users across the Facebook family of apps, Facebook likely manages billions of auctions every day, each of which consists of hundreds of relevant data points.\footnote{\url{https://www.statista.com/statistics/346167/facebook-global-dau/}, accessed on May 26, 2022.} The storage space required for all of these data would be immense.\footnote{The complexity of the task, the required investment, and the storage cost makes it very difficult to create a business case to log the required bid-request level data. Instead, it is often more cost-effective to rely on campaign-level RCTs to measure the causal effect of ads.}

This paper makes three contributions. First, we are the first to characterize the performance of a large set of experiments that are representative of the large-scale experiments that advertisers run on Facebook in the United States. We describe the results in a way that is common in the ad industry: by industry vertical and by whether the measured outcome lies in the upper, middle, or lower part of the purchase funnel. The results we describe complement the results on TV ad effects from \cite{ShapiroHitschTuchman2021} and can serve as prior distributions that digital advertisers can use for decision making.\footnote{Since our data only contain experimental outcomes for advertisers who chose to experiment, care would be necessary if attempting to generalize these results to advertisers who did not experiment. Such non-experimenting advertisers likely differ from those advertisers in our sample in a number of ways. \cite{runge_etal2020} analyzes the relative performance of advertisers who experimented on Facebook compared to advertisers who did not. \cite{runge_nair_2021} conduct a related analysis of the impact of adopting RCTs on a Facebook advertiser's subsequent advertising spending decisions.} Second, we add to the literature on whether the campaign and user-level data commonly recorded by large advertising platforms are ``good enough'' to enable non-experimental approaches to measure the causal effect of ads, or whether they prove inadequate to yield reliable estimates of advertising effects. Our results support the latter. Third, we characterize the circumstances under which a common program evaluation approach, SPSM, and a newer method at the intersection of machine learning and causal inference, DML, perform better or worse at recovering the causal effect of advertising in this scenario. We conclude that relying on non-experimental data is unlikely to succeed when ad delivery is the result of complex ranking processes unless advertising platforms fundamentally change what data they log---a change for which a business case is probably hard to make. Therefore, we believe that the exogenous variation created by RCTs will remain necessary to accurately measure ad effects. 

This paper follows a series of pioneering studies that evaluate the performance of observational methods in gauging digital advertising effectiveness. \cite{LewisRaoReiley2011} is the first paper to compare RCT estimates with results obtained using observational methods (comparing exposed versus unexposed users and regression). They faced the challenge of finding a valid control group of unexposed users: their experiment exposed 95\% of all US-based traffic to the focal ad, leading them to use a matched sample of unexposed international users. \cite{BlakeNoskoTadelis2015} documents that non-experimental measurement can lead to highly sub-optimal spending decisions for online search ads. \cite{du_et_al_2019} present a production-level system for multi-touch attribution (MTA) at JD.com using a deep learning model to estimate advertising effects. The closest paper to ours is \cite{lift1}, which analyzed the performance of certain observational methods from the program evaluation literature \citep{ImbensRubin2015} using 15 large-scale RCTs at Facebook. This paper improves upon \cite{lift1} in three ways: First, we analyze 663 experiments that were chosen to be representative of the large-scale experiments advertisers run on Facebook in the United States. Second, instead of focusing only on traditional methods from the program evaluation literature, we also use DML, a newer method at the intersection of machine learning and causal inference \citep{dml_2018, AtheyTibshiraniWager2019,deepIV2017}. Third, we use a much larger set of observable features than \cite{lift1}.

A recent paper by \cite{tunuguntla_2021} proposes a novel observational method to estimate ad effects. The approach circumvents some of the typical endogeneity problems by framing treatment as the effect of the advertiser's bid, as opposed to focusing on the endogenous outcome of exposure.\footnote{\cite{hoban_arora_2018} used similarly rich user-bid level data to estimate ad effects in an observational model, albeit without the benefit of having an RCT for comparison. A key insight in \cite{hoban_arora_2018} is that the output of targeting algorithms, such as predicted conversion probabilities, can serve as important observables to reduce the endogeneity of exposure.} The effect of an impression is calculated using a counterfactual that predicts outcomes as if the bid had been zero. The paper shows the proposed method accurately recovers the ad effect when compared to the effect obtained from an RCT. It relies on detailed bid request-level data to estimate flexible representations of user-state transition probabilities as a function of bids and the advertiser's bidding policy function. Unfortunately, these data requirements make it infeasible for many large ad platforms to apply the methodology in \cite{tunuguntla_2021} because they often do not log and retain such granular information (see the discussion above). Notably, \cite{tunuguntla_2021} solves this problem by building and bidding using his own Demand Side Platform (DSP). 

Our analysis is similar in spirit to \cite{Lalonde86} and the literature that followed (e.g., \cite{HeckmanIchimuraTodd97} and \cite{DehejiaWahba2002}). However, our source of a control group differs from that used in this other literature. \cite{Lalonde86} evaluated the effects of job training programs on labor market outcomes as part of the National Supported Work Demonstration (NSW), comparing causal effects from an experiment to those obtained using observational methods. The observational analysis compared individuals in the experimental treatment group to various control groups drawn from entirely different data sets, such as the Panel Study of Income Dynamics (PSID), which is a large, stratified random sample of the US population. 

In contrast, in our setting, the control group is composed of users assigned to the test group but who were unexposed during the campaign. The test group corresponds to a hypothetical in which an advertiser had not executed a randomized experiment, running the campaign as usual without an explicit control group. The fact that we rely on a control group from within all eligible treated users, and not a separate sample, introduces two differences relative to prior literature. First, given the design of digital advertising campaigns, there is no separate sample of users we could use to identify a control group. In the context of our hypothetical, given an advertiser's targeting criteria, all users on the platform are made eligible for ad exposure. The only pool of available control users are those who happened not to be exposed during the campaign.\footnote{Some individuals in the PSID may have wanted to join the NSW training program, had they been offered the chance. However, the NSW was geared towards disadvantaged workers lacking basic job skills. The vast majority of individuals in the PSID had no use for the NSW program. This makes it especially important to use a method, such as matching, to identify the right subset of individuals in the PSID to form a control group.} Second, although we could refer to unexposed users as ``non-compliers,'' in keeping with the literature on experiments with one-sided non-compliance \citep{angrist_imbens_rubin_1996}, these users did not fail to comply as a result of their own deliberate decisions. Instead, a combination of endogenous (Section \ref{sec:selection}) and exogenous (Section \ref{sec:obs_methods}) factors determined exposure.\footnote{Moreover, if campaign length and budget were increased, additional unexposed users might become exposed.}

This paper proceeds as follows. Section~\ref{section 2} reviews how advertising works at Facebook, how Facebook implements RCTs, and what determines advertising exposure. In Section~\ref{sec: Data} we describe how we selected our experiments, give an overview of the experiments, and describe the user-level features we use to estimate DML and SPSM. In Section~\ref{sec: Analysis of Experiments} we explain how we measure the causal effect of an advertising campaign and then present the results of our RCTs. In Section~\ref{sec: Main Results}, we introduce the two methods we use to estimate the causal effect of advertising and show the results. Section~\ref{sec: Explaining Performance} explores when our non-experimental approaches do better or worse. Section~\ref{sec: Conclusion} offers concluding remarks and suggests paths forward for academics and industry.

\section{Overview of Ad Experiments at Facebook} \label{section 2}

In this section, we describe how Facebook conducts advertising campaign experiments. Much, if not all, of the explanations here apply to advertising on Facebook, Instagram, and the Facebook (Meta) Audience Network.\footnote{\url{https://www.facebook.com/audiencenetwork/}.} We explain how these experiments work and describe the measurement challenge in advertising. This is an abbreviated version of the discussion in Section 2 of \cite{lift1}.

\subsection{Ads at Facebook} \label{sec:facebookads}
Facebook provides advertisers a number of tools and choices for designing new ad campaigns. To launch an ad campaign, an advertiser needs to make three decisions. First, the advertiser needs to choose the primary objective of their campaign. The choices for an objective include increasing awareness of their brand, improving consideration through engagement with the campaign's media, or driving conversions such as sales. Given the chosen objective, Facebook's ad platform will aim to find a broad audience of users who are likely to take the intended action and are more likely to respond positively to the ad. The second choice advertisers need to make is to refine the potential audience defined by the earlier choice of objective. For example, if an advertiser selects conversions as their objective, Facebook will find users who are more likely to convert from among the entire population on Facebook. However, an advertiser may choose to refine this population by focusing on only a particular age range, geographic location, set of interests, or previously observed behaviors. The choice of objective and target audience determine which users may potentially be served an ad from a specific advertiser. Finally, after making these choices that describe the advertiser's target audience, the advertiser chooses the ``creative'' for their ad. This involves making selecting the ad's image or video, the dimensions of the ad, and the overall design including text and other visuals.

Like most other online advertiser platforms, ads on Facebook are delivered as the result of an auction. This auction is a modified version of a second-price auction where the winning bidder pays only the minimum amount necessary to have won the auction. To balance whether the winning ad from an auction maximizes value for both users and advertisers, the final bid considered in the auction is made up of three components: (1) the bid placed by the advertiser, (2) the probability that the user in the auction will take the action consistent with the advertiser's desired objective, and (3) the quality of the ad (derived from feedback about the ad, whether people hide the ad, and by identifying potential ``low-quality'' attributes of the ad).\footnote{\url{https://www.facebook.com/business/help/430291176997542?id=561906377587030}}

Throughout this paper, we focus on ad campaigns where the advertiser is looking to drive conversions, such as purchases, signing up for a mailing list, or viewing a specific web page. In practice, these conversion events are measured through a ``conversion pixel'' which is a small piece of code provided by Facebook that advertisers add to specific pages on their website to log specific outcomes. A conversion pixel ``fires'' when the page it is on is loaded by a user, reporting information about the event back to Facebook for measurement purposes. For example, to log a purchase, an advertiser may place a pixel on an order confirmation page which would only be served and loaded if a sale was finalized. Since pixels are only attached to web pages owned by advertisers, Facebook relies on advertisers to classify the type of outcome that is measured by a corresponding pixel.\footnote{Advertisers focusing on measuring outcome events from within mobile apps may also consider building on their pixel setup through the use of the Facebook SDK (\url{https://www.facebook.com/business/help/1989760861301766?id=378777162599537}).}

\subsection{Conversion Lift} \label{sec:conversionlift}

To measure the effectiveness of a conversion-focused ad campaign, advertisers can utilize Facebook's ``Conversion Lift'' product to setup an advertising experiment.\footnote{\url{https://www.facebook.com/business/m/one-sheeters/conversion-lift}} In Conversion Lift, the ad platform randomly assigns all users in the advertiser's target audience to either a test or control group according to the advertiser's preferred proportion. In the test group, users may receive an ad from the advertiser if that ad wins the auction (we discuss why a test group user may not be exposed in the next section).

In the control group, users are guaranteed not to see an ad from the advertiser's campaign. However, this ad still participates in the auction to enable a fair comparison for the sake of ad measurement. When an ad from the advertiser running a Conversion Lift experiment wins the auction, Facebook will instead serve the user the second-place ad, which would have won if that advertiser's ad had not been running. The focal ad must remain in the auction until this last step to ensure that the correct second-place ad is shown to the control user. The result of this process is that users in the control group may be served a variety of ads from many advertisers due to the number of competitors and the diverse set of users involved. Whatever ad they are shown, it corresponds to the correct counterfactual ad that would have been displayed in the absence of the ad campaign from the focal advertiser. The Conversion Lift product is offered at no additional cost to advertisers.  Instead, the cost is borne by Facebook and consists of the difference between what Facebook charges for the highest and second-highest ranked ads for auctions in the control group. In practice this cost is small because the pool of competing advertisers is large.

Facebook defines the set of users eligible for measurement as those that (a) satisfy the targeting criteria of the advertiser (e.g., the ads should target women, age 18-49, on mobile devices) and (b) for whom the focal ad participated in at least one auction during the campaign, regardless of the outcome of that auction (i.e., independent of whether the ad was shown to the user). Those users who satisfy these criteria are known as the ``opportunity set,'' since they all at least had some positive probability of seeing the ad from the advertiser (independent of the experiment). This opportunity set is what Facebook uses to define the population for a Lift experiment, and all measurement is based on this set of users. 

While Facebook implements the same counterfactual as Google's Ghost Ads system \citep{ghostads}, the measurement sample used in the two approaches differs. Specifically, Google uses a Predicted Ghost Ads system to restrict the measurement sample to test group users who saw the ad and to control group users who are predicted to have been exposed through a simulated auction. These control users are instead shown the ad that would have been delivered after removing the focal ad from the auction. This approach yields an estimate of the Local Average Treatment Effect (LATE). In contrast, Facebook measures advertising effects using an Intent-to-Treat (ITT) approach, that compares test group users who were eligible to see the ad to control group users who were not shown the ad. Facebook can convert this effect into an ATT by scaling the ITT estimate by the share of exposed users in the test group. Facebook's design effectively corresponds to the Ghost Bids approach in \cite{ghostads}, except it is being implemented directly by the ad platform.

Facebook employs a single-user login and an identifier that persists across devices, reducing concerns about identity fragmentation \citep{coey_bailey_2016,lin_misra_2022}. Since this identifier is present for any ad exposure, ad experiments at Facebook avoid contamination between test and control groups. Thus, the structure of ad experiments at Facebook results in an unbiased measure of the effect of an ad campaign. These results measure the average treatment effect for the ad media part of the campaign on Facebook and do not generalize to media being run on other channels and are not a measure of future ad effects at a different point in time.

\subsection{Ad Exposure} \label{sec:selection}

While users in the control group are never shown an ad from a campaign running a Conversion Lift experiment, users in the test group may or may not be exposed to an ad. Exposure to an ad in the test group is not completely random---it is due to factors such as user behavior and activity, advertiser characteristics, and platform-level details about the auction.

Users must visit Facebook during a campaign to be exposed to an ad. However, users that are more likely to visit Facebook are also generally more active on the web, and are more likely to take the online action that meets the objective of the advertiser. Additionally, each time a user visits Facebook and an ad auction takes place, the diversity of competing advertisers can vary drastically depending on features such as the time of day and market conditions. While an advertiser that values a user highly will most likely be towards the top of the bid ranking, any one advertiser will not be guaranteed to win a specific user in a specific auction due to the choices of other advertisers outside of their control. Finally, modern ad delivery systems rely on a complex set of features and predictive models. After an advertiser makes audience targeting choices during campaign setup, the delivery system will continuously make updated predictions on whether a specific user is likely to take the action the advertiser's desired action. As a result, throughout the course of a campaign, the specific users that are more likely to be exposed to an ad may change.

This paper relies on this experimental setup at Facebook as it lets us measure the causal effects of an advertiser's ads through a comparison of the test and control groups, but it also enables us to leverage the one-sided compliance in the test group to mimic a setting where an advertiser chose not to run an experiment but to simply run an ad campaign on Facebook.

\section{Data} \label{sec: Data}

This section describes how we selected our experiments, gives an overview of the experiments, and describes the user-level features we use to estimate the observational models.

\subsection{Experiment Selection}

The advertising experiments analyzed in this paper were chosen to be representative of large-scale advertising experiments run in the United States on the Facebook ad platform. Ads in these experiments can appear on Facebook, Instagram or the Facebook Audience Network. These experiments cover a wide range of verticals, targeting choices, campaign objectives, conversion outcomes, sample sizes, and test/control splits. The experiments we analyze are a random subset from the set of experiments started between November 1, 2019, and March 1, 2020, and had at least one million users in the test group.\footnote{In November 2020, Facebook disclosed a bug in the experimentation platform that led to incorrect conversion metrics being reported to advertisers for about a year \citep{adexchanger_2020}. The data used for this paper was unaffected by this bug because we obtained the data from a source further upstream in the conversion measurement pipeline at Facebook.} For each experiment, we selected all outcomes with at least 5,000 conversions in the test group.\footnote{These minimums were selected by first starting with larger cutoffs during initial versions of these analyses in an effort to provide observational models the largest possible amount of data. We then continually lowered these thresholds to randomly add additional experiments while balancing overall computational resources until timing became the main constraint, such that adding additional experiments was infeasible. Note that this threshold does not necessarily select for campaigns with greater lift, since the control group may have an equally large number of organic conversions. However, it is more likely that this will select for larger campaigns overall, resulting in greater statistical power to detect a statistically significant lift. Holding campaign size fixed, this rule will select for campaigns with greater lift. } 

\subsection{Overview of Experiments and Conversion Events}

Our dataset consists of results from 563 experiments run on Facebook in the United States.\footnote{An earlier version of this paper, dated February 17, 2021, used 850 experiments that were selected using the same selection criteria. This earlier version only presented results using SPSM. Updating the analysis to include DML required us to drop some of the experiments because data retention policies at Facebook meant that the necessary individual-level data were no longer available. From an expositional perspective, we opted to present results from both models using the same sample of 563 experiments. The results for SPSM are similar when using the superset of 850 experiments.} Of these, 75 contained more than one treatment-control pair, giving us a total of 663 treatment-control pairs.\footnote{At Facebook, one treatment-control pair is called a ``cell.'' These 75 experiments are known as multi-cell tests, which advertisers might use to test different types of creatives, targeting strategies, bidding strategies, or other elements of an advertising campaign. We observe 54 experiments with two treatment-control pairs, 17 with three treatment-control pairs, and 4 with four treatment-control pairs. Please note that there is no dependency between treatment-control pairs for the same experiment. For example, if an advertiser runs two cells, no user who is assigned to one cell will be assigned to the other cell.} 
%Table~\ref{cell_distribution.table} shows the distribution of the number of treatment-control pairs conducted in each experiment.
For simplicity, we refer to each treatment-control pair as an ``experiment.'' 

\begin{figure}[t]
\center
\caption{Distribution of Experiment Characteristics}
\label{distributions.tex}
\input{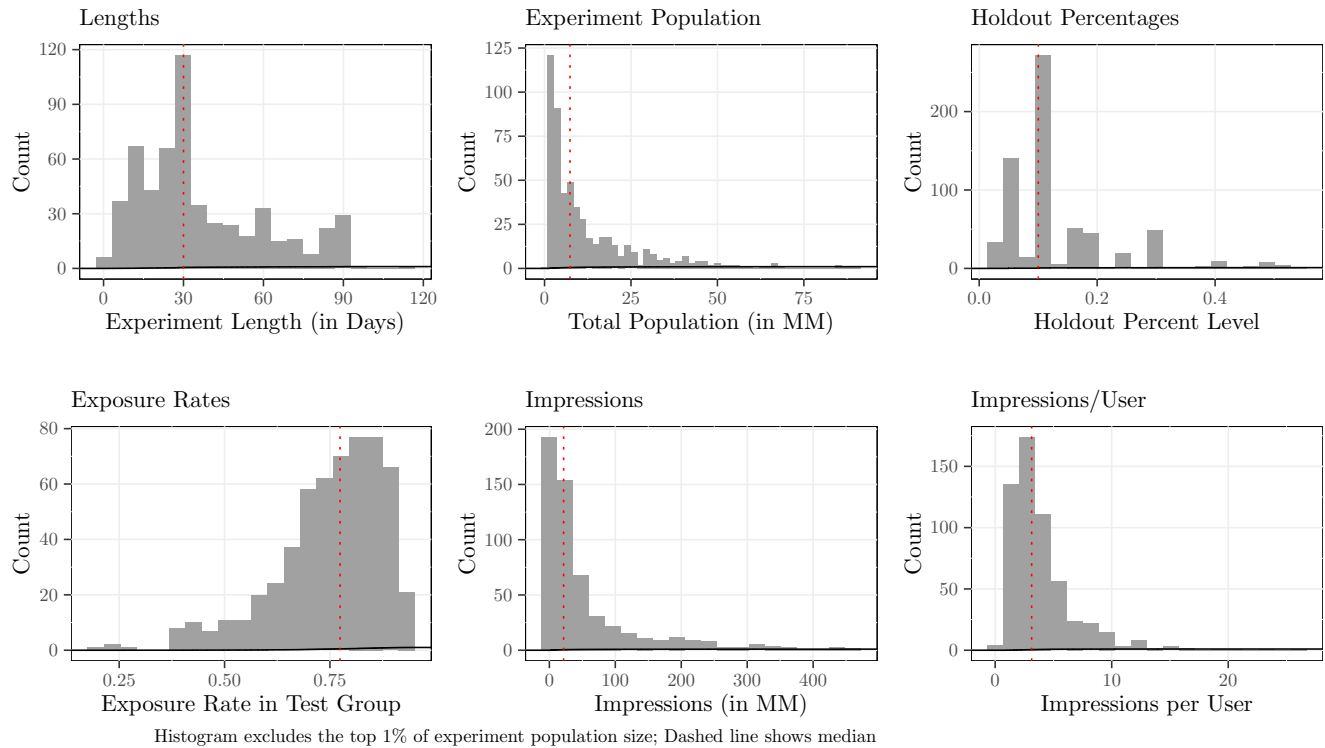}
\end{figure}

As Figure~\ref{distributions.tex} shows, experiments vary widely by length, by population size, by the fraction of users in the holdout group, by the rate at which targeted consumers were exposed, and the number of impressions. The median of experiment length is 30 days and includes 7,372,103 users across test and control groups. The median holdout percentage places 90\% of users in the test group and 10\% in the control group. For those in the test group, the median exposure percentage was 77\%, while 23\% of users were never exposed. The median of ad impressions per experiment is 22,115,390. Overall, our data set represents approximately 7.9 billion user-experiment observations with 38.4 billion ad impressions. 

We conducted a randomization check to verify that each experiment was split into test and control groups in accordance with the planned splits. For each experiment, we calculate the percentage of users in the test and control groups, and then compare this with the planned split using an exact binomial test. The distribution of p-values from this test should be approximately distributed $\textrm{Uniform}(0,1)$ if we consistently fail to reject the null hypothesis that the realized test/control split differs from the planned split. Overall, we observe that 5\% of p-values are below 0.05, 26\% are below 0.25, and 78\% are below 0.75, indicating that the experiments were properly randomized into their expected test/control splits.

%% BG - removed this on 4/14/22, absorbed into the footnote earlier.
%\input{cell_distribution.table}

Most experiments measure several different conversion outcomes, such as purchases, page views, downloads, etc. We treat all such outcomes as binary events, i.e., a user either viewed a particular webpage or they did not. Industry practitioners classify conversion outcomes by whether they occur earlier or later in a hypothetical ``purchase funnel.'' For example, page views occur early in the purchase funnel, adding items to a cart occurs later, and purchase occurs last. Our 663 experiments capture a total of 1,673 conversion events, measuring different conversion outcomes. Henceforth, we will refer to each experiment-conversion event as an ``RCT.'' We classify RCTs into ``Upper Funnel'' (601), ``Mid Funnel'' (475), and ``Lower Funnel'' (597). As we describe in Section~\ref{sec:facebookads}, outcomes are measured using ``pixels'' which advertisers choose to place on their online properties. Table~\ref{distribution_event_name.table} shows the distribution of RCTs (described by their pixel), grouped by when they occur in the purchase funnel.

\begin{table}[!htbp] \centering 
  \caption{Distribution of Conversion Events} 
  \lspace{1}
\label{distribution_event_name.table} 
\small 
\begin{tabular}{@{\extracolsep{0pt}} lccc} 
\\[-1.8ex]\hline 
\hline \\[-1.8ex] 
Pixel Name & Funnel Position & N & Percent \\ 
\hline \\[-1.8ex] 
view\_content & Upper & 410 & 24.5 \\ 
search & Upper & 121 & 7.2 \\ 
lead\_referral & Upper & 70 & 4.2 \\ 
add\_to\_cart & Mid & 266 & 15.9 \\ 
initiate\_checkout & Mid & 138 & 8.2 \\ 
add\_to\_wishlist & Mid & 34 & 2 \\ 
add\_payment\_info & Mid & 21 & 1.3 \\ 
tutorial\_completion & Mid & 16 & 1 \\ 
purchase & Lower & 409 & 24.4 \\ 
app\_activate\_launch & Lower & 97 & 5.8 \\ 
complete\_registration & Lower & 91 & 5.4 \\ 
\hline \\[-1.8ex] 
\end{tabular} 
\end{table}

Our advertisers come from many different industry verticals. Table~\ref{distribution_vertical.table} shows the distribution of RCTs by vertical.\footnote{Some smaller verticals were combined to ensure all analyses were sufficiently sized to prevent identifying individual advertisers.} E-commerce, Retail, Travel, and Entertainment/Media make up over 75\% of the RCTs we observe.

\begin{table}[!htbp] \centering 
  \caption{Conversion Events by Industry Vertical} 
  \lspace{1}
\label{distribution_vertical.table} 
\small 
\begin{tabular}{@{\extracolsep{0pt}} lcc} 
\\[-1.8ex]\hline 
\hline \\[-1.8ex] 
Industry Vertical & N & Percent \\ 
\hline \\[-1.8ex] 
E-Commerce & 504 & 30.1 \\ 
Retail & 377 & 22.5 \\ 
Financial Services/Travel & 322 & 19.2 \\ 
Entertainment/Media & 145 & 8.7 \\ 
Tech/Telecom & 124 & 7.4 \\ 
Consumer Packaged Goods & 105 & 6.3 \\ 
Other & 96 & 5.7 \\ 
\hline \\[-1.8ex] 
\end{tabular} 
\end{table}

% Advertisers often measure outcomes across the purchase funnel, they typically only measure one specific outcome event per position. Since we will focus on analyzing outcomes at the funnel level, the RCTs we analyze are independent (i.e., within a specific funnel position, RCTs consist of independent populations, with a few exceptions we will point out in the analysis). Therefore, we define RCTs as the unit of observation for the remainder of the paper.
Advertisers measure outcomes across the purchase funnel. Since we define RCTs as an experiment (e.g., a specific treatment-control pair) and conversion event, we use RCTs as the unit of observation for the remainder of the paper and analyze outcomes by funnel.

\subsection{User Features}\label{user features}
For each user in an experiment, we observe a large set of variables logged before users are potentially exposed to an ad in the campaign. We group these features as describing a dense set of descriptive user features, a sparse set of user interest features, estimated action rates, and prior campaign-related outcome activity. These features play a significant role in serving ads as they directly contribute to determining the winner of an ad auction and describing whether a user is likely to convert for any given experiment. Here is a description of each feature group:

\begin{enumerate}
   \item \textbf{Dense features.} User characteristics such as gender, age, household size, as well as Facebook-specific attributes describing the age of the user's account, number of posts, friends, Likes, and comments, devices used to access Facebook, and measures of activity such as ad impressions, clicks and conversions across the Facebook family of apps. Although some of these variables are effectively time invariant (e.g., gender, age), measures of activity are logged in rolling windows of either 84 or 28 days, and so these vary across user-experiment combinations. We log these characteristics on the Sunday before the start of each advertising experiment.
   \item \textbf{Sparse features.} Facebook allows users to express interests in a large number of different topics. For example, users can express interest in hobbies such as cooking, watching certain genres of movies, listening to certain kinds of music, playing different sports or video games, or being interested in technology, science, the outdoors, or traveling. We log sparse features on the Sunday before the start of each advertising experiment.
   \item \textbf{Estimated action rates.} As described in Section~\ref{sec:facebookads}, when a user is first considered as a candidate for exposure in an ad auction for a particular advertiser, Facebook estimates the probability that showing the ad to this user will lead to a desired advertiser outcome.\footnote{These are similar to the ``targeting score'' discussed in \cite{hoban_arora_2018}.} We log these features the first time the ad auction considers a user for a given advertising experiment.
   \item \textbf{Prior campaign outcomes.} To measure the results of advertising campaigns, advertisers use conversion pixels (see Section~\ref{sec:facebookads}) to log user outcomes. We measure up to a month of prior conversion data for each outcome event, depending on when the advertiser installed a conversion pixel. \end{enumerate}

The second group consists of thousands of features, whereas the remaining feature groups comprise roughly 500 descriptive variables. Due to the differences in treated and untreated groups inherent in an observational analysis setup, we utilize these features to create a balanced set of exposed and unexposed users, specifically for understanding treatment status, to satisfy unconfoundedness as described earlier. These features are also used for purely predictive aspects of estimation, namely when describing conversion outcomes for users. While some of these features are specific to the Facebook platform, many other digital services would have analogs that describe similar sets of features as those we describe.

\section{Analysis of Experiments} \label{sec: Analysis of Experiments}

In this section, we first explain how we measure the causal effect of an advertising campaign and then present the results using our 1,673 RCTs. 

\subsection{Methods}

To explain our measurement approach, we make use of the potential outcomes notation. In this subsection we summarize the exposition in \cite{lift1}, which in turn used material in \cite{Imbens2004}, \cite{Imbens_wooldridge2009}, and \cite{ImbensRubin2015}. All variables below are specific to an RCT, and so we do not include such a subscript.

Each experiment contains $N$ individuals who are randomly assigned to test or control conditions through $Z_{i} = \{0,1\}$. Exposure to ads is given by $W_{i}(Z_{i}) = \{0,1\}$. Users assigned to the control condition are never exposed to any ads from the experiment, $W_{i}(Z_{i} = 0) = 0$. However, exposure is endogenous outcome among users assigned to the test group, such that $W_{i}(Z_{i} = 1) = \{0,1\}$ (i.e., there is one-sided non-compliance). We observe a set of features $X_{i} \in \mathbb{X} \subset \mathbb{R}^{P}$ for each user that are unaffected by the experiment. The potential outcomes are $Y_{i}(Z_{i}, W_{i}(Z_{i})) = \{0,1\}$. Given a realization of the assignment, and the subsequent realization of the endogenous exposure variable, we observe the triple $Z_{i}$, $W_{i} = W_{i}(Z_{i})$, and $Y_{i} = Y_{i}(Z_{i}, W_{i})$.

%Given an assignment $Z_{i}$ and a treatment $W_{i}(Z_{i})$, the potential outcomes are $Y_{i}(Z_{i}, W_{i}(Z_{i})) = \{0,1\}$. Under one-sided noncompliance, the observed outcome is
%\begin{align}
%Y_{i}^{obs} = Y_{i}(Z_{i},W_{i}^{obs}) = Y_{i}(Z_{i},W_{i}(Z_{i})) = \left\{ \begin{array}{cc}
%Y_{i}(0,0), & \mbox{ if } Z_{i} = 0, W_{i}^{obs}= 0 \\
%Y_{i}(1,0), & \mbox{ if } Z_{i} = 1, W_{i}^{obs} = 0 \\
%Y_{i}(1,1), & \mbox{ if } Z_{i} = 1, W_{i}^{obs} = 1 \end{array} \right .
%\end{align}
%We designate the observed values $Y_{i}^{obs}$ and $W_{i}^{obs}$ to help distinguish them from their potential outcomes.

As a first step, the intent-to-treat (ITT) effect compares outcomes across random assignment status:
\begin{align}
\mbox{ITT} & = \E \left[Y(1,W(1)) -  Y(0,W(0)) \right] \ ,
\end{align}
with the sample analog being
\begin{align}
\widehat{\mbox{ITT}} & = \frac{1}{N} \sum_{i=1}^{N} \left( Y_{i}(1,W_{i}) - Y_{i}(0,W_{i}) \right) \ .
\end{align}
This calculation rests on the Stable Unit Treatment Value Assumption (SUTVA) \citep{rubin1978}, which requires that a user only receive one version of the treatment and that a user's treatment assignment does not interfere with another user's potential outcomes. The ad experiments on Facebook should satisfy both conditions. The platform's single-user login design helps ensure it shows the right ad to the right user. Although the general form of interference is untestable, we would not expect significant ``spillover'' effects in the context of these online ad experiments.

However, the observational models we use do not produce ITT estimates because they lack experimental control groups. To compare our RCT estimates with those obtained from observational methods, we instead estimate the average treatment effect on the treated (ATT),
\begin{align}
\att{} \equiv \mbox{ATT} &  = \E \left[Y(1,W(1)) -  Y(0,W(0)) | W(1) = 1 \right] \ .
\end{align}

To estimate the ATT, we rely on the following exclusion restriction: $Y_{i}(0, W)  = Y_{i}(1, W)$, for any $W$. This assumption requires that random assignment only affects a user's outcome through receipt of the treatment. Using this assumption, we estimate the ATT using two-stage least squares (2SLS) with assignment $Z$ as an instrument for endogenous exposure $W$ \citep{ImbensAngrist94}. The estimate we obtain, $\hat{\tau}$, can be interpreted as the average effect among exposed users.\footnote{When we interpret the ATT, it is always conditional on the entire treatment (e.g., a specific ad delivered on a particular day and time) and who is targeted with the treatment. In the context of online advertising, the ``entire treatment'' includes the advertising platform and its ad-optimization system.}
 
\subsection{Results}

\begin{figure}
 	\centering
	\caption{ATTs across all RCTs}
	\label{distribution_ATT.tex}
	\input{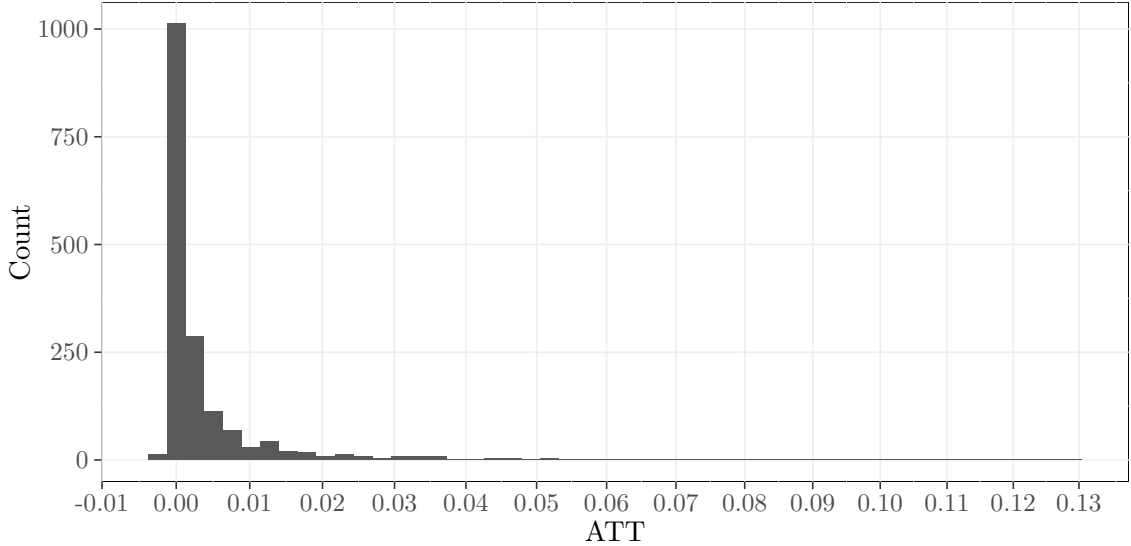}
\end{figure}

We begin by reporting the estimated ATTs across all 1,673 RCTs. As Figure~\ref{distribution_ATT.tex} shows, most ATTs are below 0.01, while some can be as high as 0.13. The median ATTs from upper to lower funnel outcomes are 0.003, 0.001, and 0.0002, respectively (see Appendix Table A-1 for deciles by funnel level). We obtain the standard error of the ATT through the 2SLS regression. Using this estimate, 1,170 of 1,673 RCTs, or 70\%, are statistically significant using a two-tailed t-test at the $\alpha=0.05$ level. Among these, 1,160 RCTs have positive and significant ATTs.  

However, ATTs are difficult to interpret since they contain no information on whether the ATT is ``small'' or ``large.'' Hence, to more easily interpret outcomes across RCTs, we report most results in terms of \textit{lift}, the incremental conversion rate among treated users expressed as a percentage,

\begin{align}
\ell &=\frac{\mbox{Conversion rate due to ads in the treated group}}{\mbox{Conversion rate of the treated group if they had \textit{not} been treated}} \nonumber \\
	& = \frac{\att{}}{\E [Y | Z = 1, W =1]  - \att{} }. \label{eq:lift_att}
\end{align}
The denominator is the estimated conversion rate of the treated group if they had not been treated. Reporting the lift facilitates the comparison of advertising effects across RCTs because it normalizes the results according to the treated group's baseline conversion rate, which can vary significantly with experiment characteristics (e.g., advertiser's identity, the outcome of interest).

\begin{figure}
 	\centering
	\caption{Lifts across all RCTs}
	\label{distribution_Lift.tex}
	\input{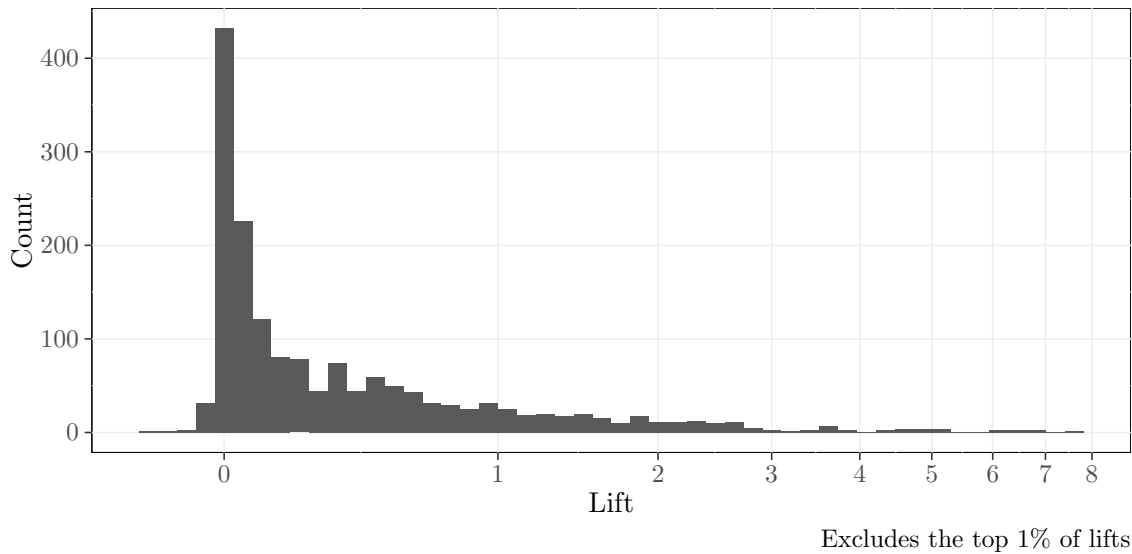}
\end{figure}

\begin{figure}
 	\centering
	\caption{Lifts by Purchase Funnel Position}
	\label{distribution_Lift_boxplots.tex}
	\input{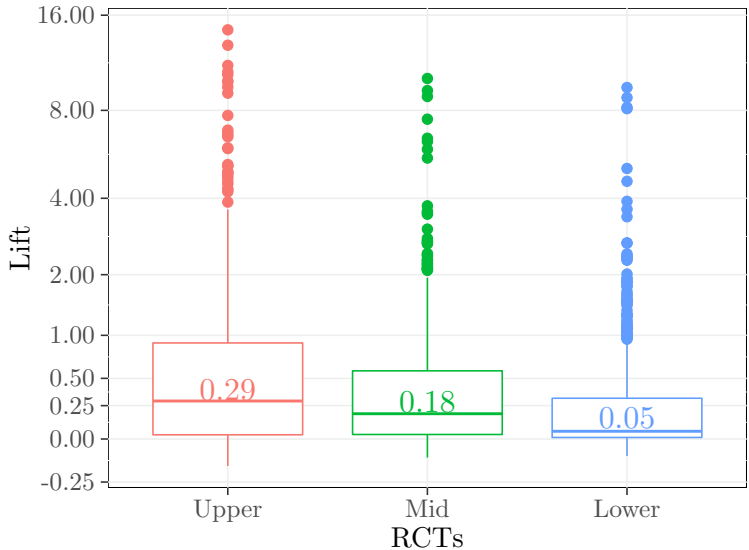}
\end{figure}

Figure~\ref{distribution_Lift.tex} shows the distribution of lifts across all RCTs. The average lift is 52\%, while the median lift is 9\%. \cite{JohnsonLewisNubbemeyer2017} report a similar median lift estimate in their analysis of 432 experiments on the Google Display Network.\footnote{As an additional point of comparison, we calculate Cohen's $d$ \citep{cohen_1977} for each RCT, dividing the ITT effect by the pooled standard error of the estimate. The median values for $d$ are 0.0036, 0.0127, and 0.0279 for lower, middle, and upper funnel outcomes, respectively. See Table 3 of \cite{johnson2022} for Cohen's $d$ values for other display advertising experiments.} Figure~\ref{distribution_Lift.tex}, however, masks differences between event funnel outcomes. As Figure~\ref{distribution_Lift_boxplots.tex} shows, the median lower funnel outcome is smaller than the median lift of mid-funnel outcomes while upper funnel outcomes are higher still.

Variance across RCTs can vary substantially. Figure~\ref{distribution_Lift_across_funnels.tex} displays lifts by the conversion outcome's position along the purchase funnel together with 95 percentile bootstrapped confidence intervals. We use the standard errors obtained from these bootstraps to conduct inference on the lift estimates. The plot indicates whether the lift is statistically different from zero at the 5\% level (two-tailed t-test using bootstrapped standard errors). We find that 75.8\% of upper-funnel RCTs have lifts that are statistically different from zero. For mid-funnel RCTs, this number is 73.7\%, while 59.6\% of lower-funnel RCTs are statistically different from zero.

\begin{figure}
 	\centering
	\caption{Lifts by Purchase Funnel Position}
	\label{distribution_Lift_across_funnels.tex}
	\input{distribution_Lift_across_funnels.tex}
\end{figure}

\begin{figure}
 	\centering
	\caption{Detectable Lift}
	\label{Lift_power.tex}
	\input{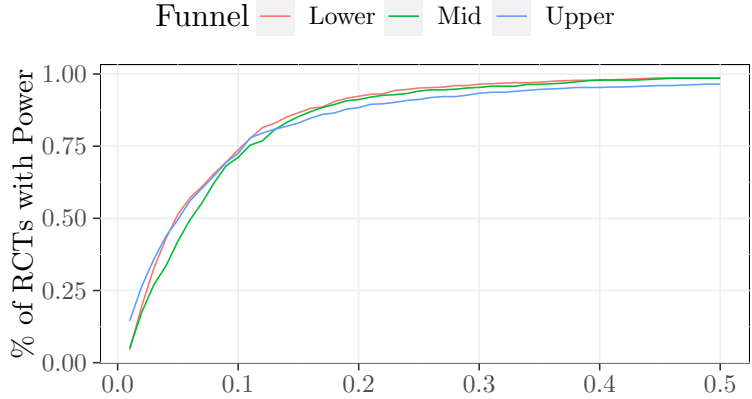}
\end{figure}

Next, to interpret the result regarding statistical significance, we want to know whether the RCTs we observe were adequately powered. Figure~\ref{Lift_power.tex} shows the proportion of RCTs that had a 50\% ex-ante power to detect a given lift at the 5\% significance level.\footnote{We follow the procedure suggested by \cite{GelmanHill2007} and implemented in \cite{ShapiroHitschTuchman2021}, namely to identify the proportion of RCTs for which the standard error of the lift estimate is less or equal to detectable lift divided by 1.96.} Across purchase funnel categories, 75\% of RCTs were powered to detect a lift of 10\%. 85-90\% of RCTs were powered to detect a lift of 20\%. Given that the lift in mid and upper-funnel RCTs is typically higher than for lower-funnel RCTs, this explains why the fraction of insignificant lift estimates is higher for lower-funnel RCTs.

\begin{figure}
 	\centering
	\caption{Distribution of Lifts by Industry Vertical}
	\label{Lift_vertical.tex}
	\input{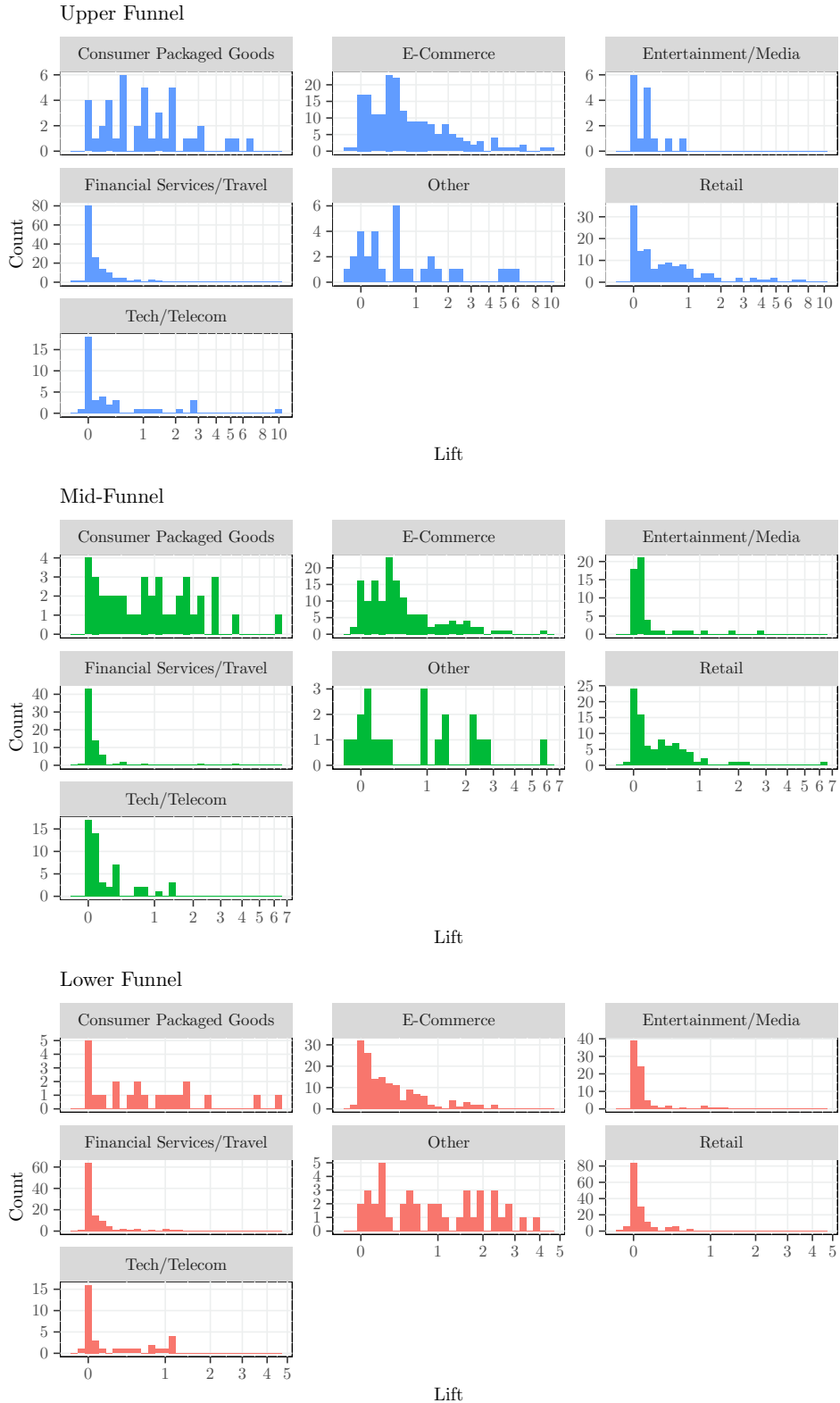}
\end{figure}

Lifts vary widely by industry vertical. Figure~\ref{Lift_vertical.tex} shows the distribution of lifts for the seven industry vertical groupings we introduced in Table~\ref{distribution_vertical.table}, separated by purchase funnel position.

\section{Observational approaches} \label{sec: Main Results}

The following thought experiment motivates the analysis in this section. Rather than conducting an RCT, an advertiser implemented an ad campaign against a target audience without an explicit no-ad control group, thus requiring the application of an observational method to measure the impact of the ad. In this setting, all users who satisfied the targeting criteria were made eligible to see the ads. Due to the endogeneity of ad exposure (see Section~\ref{sec:selection}), a non-random subset of eligible users are exposed. Comparing outcomes between exposed and unexposed users is likely to produce a biased estimate of the treatment effect due to systematic differences in the users. 

To properly estimate the causal effect, we apply stratified propensity score matching (SPSM) and double/debiased machine learning (DML), which use a large set of user-level features to adjust for differences between the exposed and unexposed groups. We discuss the key assumptions underlying each method and explain how ad exposure contains a degree of quasi-randomness that we leverage for estimation.

\subsection{Methods} \label{sec:obs_methods}

In the observational analysis that follows, we focus on the test group ($Z_i = 1$) exclusively and ignore the control group ($Z_i = 0$). With a slight abuse of notation, we redefine potential outcomes as
\begin{align}
Y_{i}(W_{i}) & \equiv Y_{i}(Z_{i} = 1 , W_{i}) \ .
\end{align}
Thus, having restricted our analysis to the test group, for each user we observe the triple $(Y_{i}, W_{i}, X_{i})$. We view this modified version of our data as corresponding to the (hypothetical) observational sample that an advertiser would use in the absence of a randomized experiment. The ATT obtained using observational method $m$ is
\begin{align}
\att{m} = \E \left[Y(1) - Y(0) | W = 1 \right] \ .  
\end{align}
and the lift is
\begin{align}
\ell^{m} & = \frac{\att{m}}{\E [Y | W =1]  - \att{m} } \ .
\end{align}

If treatment status $W_{i}$ were random and independent of $X_{i}$, comparing the conversion rates of exposed and unexposed users would provide an estimate of the campaign's causal effect. Going forward we will refer to this as the ``exposed-unexposed'' baseline. This ATT effect is
\begin{align}
\att{\eu} &= \E [ Y(1)] - \E [Y(0)] \label{eq:exp_unexp} \ ,
\end{align}
with a corresponding lift of
\begin{align*}
\ell^{\eu} & = \frac{\att{\eu}}{\E [Y(0)]} \ .
\end{align*}
This estimate serves as a useful baseline of comparison with SPSM and DML because it has not been corrected to reflect any self-selection into treatment.

However, treatment status $W_{i}$ usually is \emph{not} random and independent of $X_{i}$. Ignoring this fact will lead to bias in the resulting causal effects. In addition to SUTVA, observational models attempt to correct for this bias under several assumptions. One assumption is unconfoundedness,
\begin{align}
 \left(Y_{i}(0), Y_{i}(1)\right) \ci  W_{i}  \ \ | \ \ X_{i}  \label{eq:obs_unconf} \ ,
 \end{align}
 such that potential outcomes are independent of treatment status, conditional on $X_{i}$. This (untestable) assumption implies there are no unobserved individual characteristics that are associated with the treatment and potential outcomes. If this assumption holds, it will be because of the advertising's platform rich set of covariates. The second assumption is overlap,
 \begin{equation*}
0 < \Pr(W_{i} = 1 | X_{i} ) < 1, \ \ \forall X_{i} \in \mathbb{X} \ .
 \end{equation*}
 which requires the probability of treatment to be bounded away from zero and one for all values of user characteristics. Both uncounfoundedness and the overlap assumption are necessary to apply SPSM and DML.\footnote{For DML, in \cite{dml_2018}, see section 5.1 regarding unconfoundedness and Assumption 5.1(c) for overlap.} 
 
A possible concern is that the allocation rule used in online ad auctions may leave no room for randomness, thereby violating the overlap assumption. In reality, a user's ad exposure is subject to randomness, even under a deterministic auction process (see also  \cite{lift1}, Section 6.1, for detailed discussion). Using Facebook as an example, we discuss how randomness can be introduced in three different phases that determine whether a user is exposed to an ad:

\emph{Phase 1:} Whether and how many times a user visits Facebook during the period of the campaign is a random outcome from the perspective of the platform (and researchers). For example, if a user happens to be backcountry hiking with only intermittent Internet access, the user may have limited opportunities to be exposed to the focal advertiser's ads, even if this user is identical to other users on all characteristics Facebook relies on to target ads.

\emph{Phase 2:} Conditional on a user being on Facebook, the page will request bids for a certain number of ad slots. Whether the focal advertiser wins any of these ad slots, resulting in a (potential) ad exposure, is partly a random outcome, even though the auction allocation rule is deterministic. The reason is that these deterministic allocation rules are themselves functions of random variables. For example, whether that advertiser has exhausted their daily budget, which in turn depends on what users showed up on Facebook on that day before the specific auction. More generally, budget pacing rules induce the probabilistic inclusion of advertisers in an auction, providing a helpful source of quasi-experimental variation to estimate LATE's \citep{gui_nair_niu_2022}.

\emph{Phase 3:} Conditional on the focal advertiser winning an ad slot for a particular user, the user may not actually be exposed to the ad. For example, when a user loads their Facebook News Feed, Facebook may fill multiple ad slots with the results of different ad auctions (to minimize load times for users while they are scrolling). Whether an ad is assigned to slot two or seven is quasi-random, partly determined by random outcomes described in Phase 2 (the actions of other users and advertisers). Furthermore, whether a user scrolls far enough to see the ad is determined by quasi-random factors such as how fast they scroll or distractions in their local environment.

Thus, with these sources of randomness, even if the auction allocation rule is deterministic, identical users facing identical circumstances, except for random idiosyncratic shocks, will differ in their exposure status, therefore fulfilling the overlap assumption.

\subsubsection{Stratified Propensity Score Matching (SPSM)}\label{SPSM model}

The first method we use to address the non-randomness of treatment is propensity score matching \citep{DehejiaWahba2002,stuart2010}. The propensity score, $e(X_{i})$, is the conditional probability of treatment given features $X_{i}$,
\begin{align}
e(X_{i}) & \equiv \Pr(W_{i}=1 | X_{i} = x) \ .
\end{align}
Under strong ignorability, \cite{rosenbaum_rubin83} establish that treatment assignment and the potential outcomes are independent, conditional on the propensity score,
\begin{align}
\left(Y_{i}(0), Y_{i}(1)\right) \ci  W_{i} \ \ | \ \ e(X_{i}) \ .
\end{align}
This result shows that the bias from selection can be eliminated by adjusting for the propensity score.
 
In standard propensity score matching, we find the one (or more) unexposed users with the closest propensity score to each exposed user to estimate the treatment effect. Since this is very computationally burdensome, instead, we stratify on the propensity score: After estimating the propensity score, $\hat{e}(X_{i})$, we divide the sample into strata such that within each stratum, the estimated propensity scores are approximately constant.  This method, known as stratified propensity score matching (SPSM), scales well and achieves good feature balance without an over-reliance on extrapolation \citep{ImbensRubin2015}. \cite{lift1} found that propensity score matching and stratification on the propensity score produced similar results.

We first partition the estimated propensity scores into $J$ intervals of $[b_{j-1}, b_{j})$, for $j = 1, \ldots, J$. Let $B_{ij}$ be an indicator that user $i$ is contained in stratum $j$,
\begin{align}\label{eq:strat start}
B_{ij} & = \mathbbm{1} \cdot \left\{ b_{j-1} < \hat{e}(X_{i}) \leq b_{j}\right\}
\end{align}
Each stratum contains $N_{wj} = \sum_{i=1}^{N} \mathbbm{1} \cdot \{W_{i} = w\} B_{ij}$ observations with treatment $w$. The ATT within a stratum is estimated as
\begin{align}
\attest{spsm}_{j} & = \frac{1}{N_{1j}} \sum_{i=1}^{N} W_{i} B_{ij}Y_{i} - \frac{1}{N_{0j}} \sum_{i=1}^{N} (1-W_{i})B_{ij}Y_{i} \ .
\end{align}
The overall ATT is the weighted average of the within-strata estimates, with weights corresponding to the fraction of treated users in the stratum relative to all treated users,
\begin{align}
\attest{spsm} & = \sum_{j=1}^{J} \frac{N_{1j}}{N_{1}} \cdot \attest{spsm}_{j} \ , \label{eq:strat_att}
\end{align}
where $N_{1}$ is the number of users in the treated group. We split the sample into 100 equally spaced strata of 1\%.\footnote{We explored the approach proposed in \cite{ImbensRubin2015}, which uses the variation in the propensity scores to determine the number of strata and their boundaries, which was used in \cite{lift1}. In the present sample of RCTs, we did not find that it made a meaningful difference in our estimates while being computationally much more demanding.}

The variance of the estimator is
\begin{align}
\hat{V}_{wj} & = \frac{{S}_{wj}^{2}}{N_{wj}}, \quad \quad \textrm{where} \ \ {S}_{wj}^{2} = \frac{1}{N_{wj}} \sum_{i:B_{ij}=1,W_{i}=w} \left(Y_{i} - \bar{Y}_{wj}\right)^{2}  \\
\hat{V}(\attest{spsm})  & = \sum_{j=1}^{J} \left(\hat{V}_{0j} + \hat{V}_{1j}\right)^{2} \cdot \left( \frac{N_{1j}}{N{1}} \right)^{2} \ .
\end{align}

\subsubsection{Double/Debiased Machine Learning (DML)}\label{DML model}

In the past few years, the machine learning community has made vast improvements to predictive modeling procedures with new statistical methods and advances in computational hardware. Given the focus of these models on making accurate predictions, they are trained on data sets for which the true answer is known for a set of records and are then applied to new, unseen data. However, in causal inference settings, where the goal is not simply predictive power and where we will never observe true outcomes for any individual record, a direct application of machine learning methods to estimate causal effects can lead to invalid, biased, results. 

In recent years, new work had aimed to combine the advantages of machine learning with the causal inference goals of traditional econometrics. Specifically, new literature has addressed the main reasons why predictive models may struggle with causal inference, namely the bias that arises from regularization and overfitting. The double/debiased machine learning (DML) approach introduced by \cite{dml_2018} corrects for both of these sources of bias by using orthogonalization to account for the bias introduced by regularization and by implementing cross-fitting to remove bias introduced by overfitting. Double machine learning methods build on common econometric approaches by combining the benefits of cutting-edge machine learning with causal inference methods such as propensity score matching.

Let outcomes and treatments be given by
\begin{align}
Y_{i} & = g(W_{i}, X_{i}) + u_{i} \\
W_{i} & = e(X_{i}) + \nu_{i} \ ,
\end{align}
where we assume $E[u | X,W] = 0$ and $E[\nu | X] = 0$. Since treatment $W_i$ and user characteristics $X_i$ enter into the nonlinear function $g(\cdot)$, we allow for possibly complex heterogeneous effects of treatment on outcomes. Thus, the ATT is
\begin{align}
\tau^{dml} = E[g(1,X) - g(0,X) | W = 1].
\end{align}

To orthogonalize this setup, we can partial out the effect of $X$ on $W$ in three steps: (1) directly predicting $W$ based on $X$, (2) directly predicting $Y$ based on $X$, and (3) by regressing the residuals from step two on the residuals from step 1. In practice, we do this using the score function:
\begin{align}
\psi(X; \tau^{dml}, \eta) = \frac{W(Y - g(W, X))}{N_{1}} - \frac{e(X)(1 - W)(Y - g(W, X))}{N_{1}(1 - e(X))} - \frac{W \tau^{dml}}{N_{1}}
\end{align}
where $\eta = (g, e)$ is a nuisance parameter and $N_{1}$ is again the proportion of treated users. If we set the expectation of this function to zero, $\sum_{i=1}^NE(\psi) = 0$, we can obtain an estimate $\hat{\tau}^{dml}$.  

We use cross-fitting to account for the bias introduced by overfitting. We start by randomly partitioning our data into $K$ subsets. One partition, $k$, is held out and we fit models for $W$ and $Y$ on the remaining subsets. We use these models to estimate $\tau_{k}^{dml}$ in the held-out partition. We continue for the remaining $K-1$ partitions, holding one out and modeling on the remaining subsets, until we have an estimate of $\tau_{k}^{dml}$ for each partition. To get our final estimate, we average across the partition estimates, 
\begin{align}
\hat{\tau}^{dml} = \frac{1}{K}\sum_{k = 1}^{K} \hat{\tau}_{k}^{dml}.
\end{align}
An estimate of the variance of $\tau^{dml}$ can be calculated as
\begin{align}
	\hat{\sigma}^2 & = J^{-2} \frac{1}{N} \sum_{k=1}^{K} \sum_{i \in I_k} [\psi(X_i;  \hat{\tau}_{k}^{dml}, \hat{\eta})]^2 \\
	J & = \frac{1}{N} \sum_{k=1}^{K} \sum_{i \in I_k} \psi(X_i;  \hat{\eta}) \ ,
\end{align}
where $\hat{\eta} = (\hat{g}, \hat{e})$ indicates evaluating these expressions using the estimated functions.

\subsubsection{Estimation} \label{sec: Estimation}

Both SPSM and DML require propensity score models. We estimate the propensity score using a deep learning model due to its ability to leverage both large amounts of dense and sparse features and its ability to parallelize model training across a cluster of computers \citep{DLRM2019}. Deep learning models are characterized by a set of hyperparameters that include the network topology, dropout rate, learning rate, and the number of passes through the training data.

We build a propensity score model using the four groups of features described in Section~\ref{user features} separately for each of the 1,673 RCTs. Additionally, because hyperparameter sweeping requires training a large number of models, we determined a set of optimal hyperparameters by sweeping over a random subset of our propensity score models. We used these hyperparameters across all RCTs.

We obtain propensity scores for each user in our RCTs by training each propensity model using three-fold cross-validation. We use two folds to train a model for estimating propensity scores for the remaining fold. This process takes place three times so that each fold is used two times for training and once for estimation.

The DML model also requires an outcome model ($g(\cdot)$). Similar to our approach for the propensity score models, we trained a deep learning model using three-fold cross-validation to predict whether an individual converts. For these outcome models we used the same set of features described in Section~\ref{user features}, as well as the treatment status for each user. We selected hyperparameters by sweeping over a random subset of models and selecting the best set to use across all RCTs. In total, we trained 5,019 propensity score models and 5,019 outcome models for the 1,673 RCTs in our data. Given the size of the RCTs and the number of features, training this number of models required significant cluster computing resources.

\subsection{Model Fit}

\begin{figure}
 	\centering
	\caption{Distribution of AUCs in Propensity (SPSM and DML) and Outcome Model (DML only)}
	\label{distribution_auc.tex}
	\input{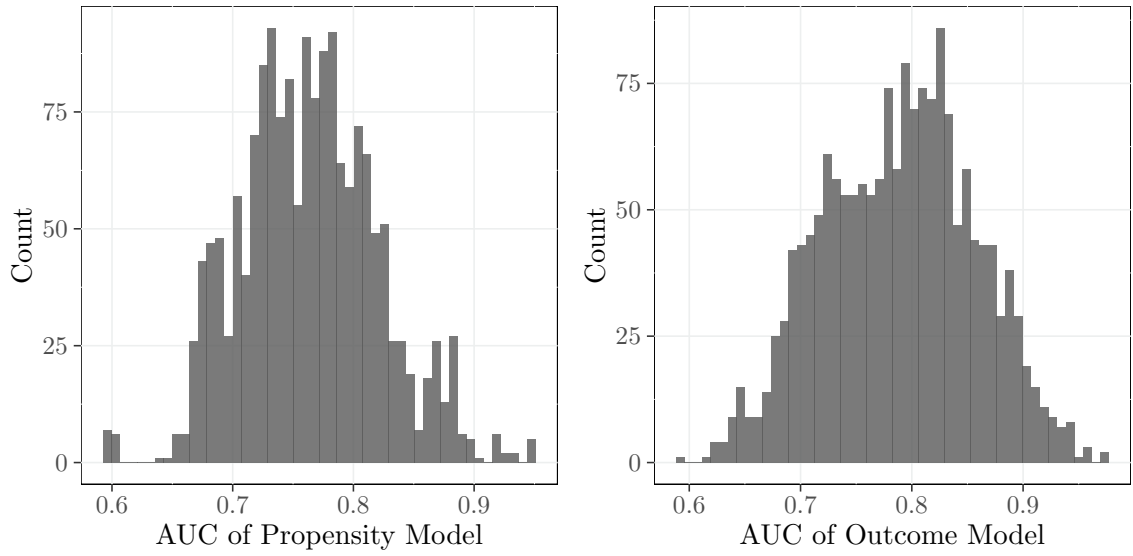}
\end{figure}

We summarize the fit of our propensity score and outcome models in Figure~\ref{distribution_auc.tex}. We see a wide distribution of model performance with an ``area under the ROC curve'' (AUC) between 0.6 and 0.95 with a median of 0.76. This variation is expected given that the RCTs in our sample represent a number of different industry verticals, outcome funnel positions, population sizes, and ad campaign targeting and delivery settings. The AUCs of the outcome model also show a wide distribution with AUCs between 0.6 and 0.98 with a median of 0.79. 

\begin{figure}
 	\centering
	\caption{Scatterplot of Propensity and Outcome Model AUCs}
	\label{scatter_auc.tex}
	\input{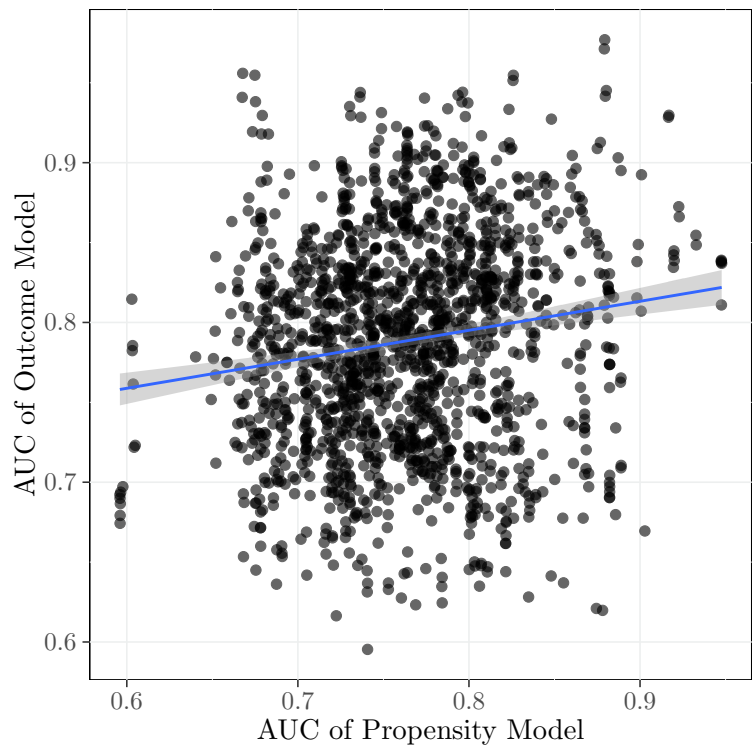}
\end{figure}

The fit of the propensity score and outcome models are not highly correlated. Figure~\ref{scatter_auc.tex} shows a scatterplot of the AUCs of the two models for each RCT. This should not be surprising given that these models predict very different outcomes. The propensity model predicts treatment assignment and---under the identifying assumption for SPSM---should not perfectly predict. In contrast, the outcome model predicts conversions and performs better when the fit is higher. 

\subsection{Results}

We begin reporting our results with a simple count of the number of RCTs where SPSM and DML results are statistically different or indistinguishable from RCT results. Table~\ref{performance_overview_table.table} shows the results for all funnels (top 2 rows), and by funnel (next 6 rows). Each comparison distinguishes whether the RCT lift was statistically significant at the 5\% level, again relying on the bootstrapped standard errors. Columns 3 and 4 show how often the difference between SPSM lift and RCT lift was statistically indistinguishable from zero, or different. Column 5 summarizes the percentage of RCTs for which SPSM and RCT results were statistically different from each other. Columns 6-9 repeat the comparison for DML. 

We reject the hypothesis that SPSM and RCT lifts are equal for 82\% of the RCTs with insignificant RCT lifts and for 91\% of RCTs with significant RCT lifts. For DML, the equivalent percentages are 77 and 75\%. Results do not differ much by funnel position. Therefore, our analysis suggests that neither SPSM not DML do a good job at recovering RCT lifts.     

\begin{table}[!htbp] \centering 
\caption{Number of RCTs where SPSM and DML results are \\statistically indistinguishable or different from the RCT results} 
\lspace{1}
\label{performance_overview_table.table} 
\small 
\begin{tabular}{@{\extracolsep{0pt}} lc|ccc|ccc} 
\\ \hline\hline
\multicolumn{2}{c|}{} & \multicolumn{3}{c|}{SPSM vs. RCT} &  \multicolumn{3}{c}{DML vs. RCT} \\ 
Funnel & RCT p-value & p$>$0.05 & p$\leq$0.05 & \% Sign.  & p$>$0.05 & p$\leq$0.05& \% Sign.  \\ \hline\hline
All & $>$0.05& 80 & 375 & 82\%&105 & 350 & 77\%\\ 
 & $\leq$0.05 & 111 & 1107 & 91\%&310 & 908 & 75\%\\ \hline \hline
Upper & $>$0.05 & 20 & 93 & 82\%&28 & 85 & 75\%\\ 
 & $\leq$0.05 & 33 & 455 & 93\%&102 & 386 & 79\%\\ \hline
Mid & $>$0.05 & 24 & 96 & 80\%&26 & 94 & 78\%\\ 
 & $\leq$0.05 & 19 & 336 & 95\%&108 & 247 & 70\%\\ \hline
Lower & $>$0.05 & 36 & 186 & 84\%&51 & 171 & 77\%\\ 
 & $\leq$0.05 & 59 & 316 & 84\%&100 & 275 & 73\%\\ \hline\hline
\end{tabular} 
\end{table}

\begin{figure}
 	\centering
	\caption{Comparison of RCT Lifts with Lifts Estimated using SPSM and DML}
	\label{Lift_RCT_SPSM_DML.tex}
	\input{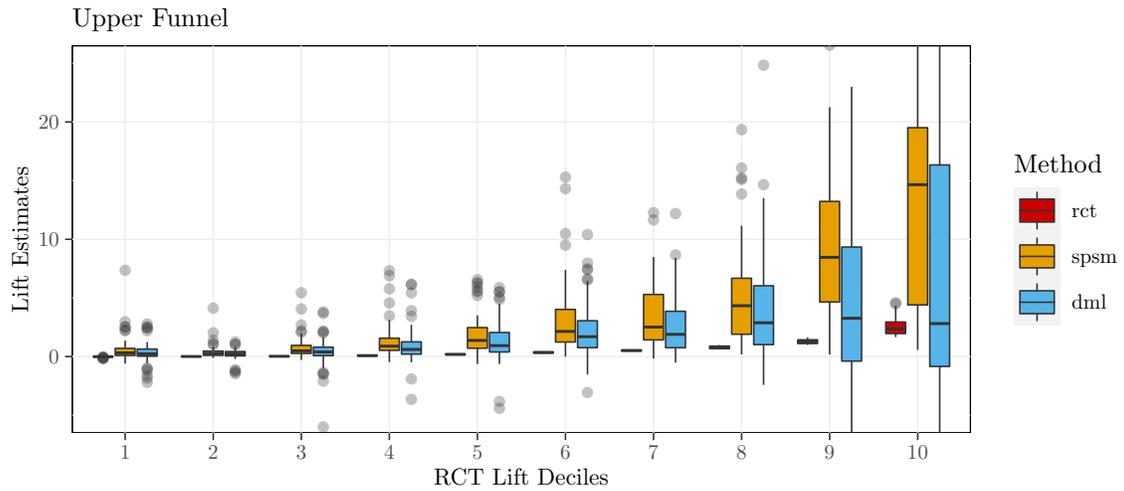}
\end{figure}

Next, we visually depict SPSM and DML lifts in Figure~\ref{Lift_RCT_SPSM_DML.tex}. To do so, for each purchase funnel position, we split all RCTs into lift deciles where decile 1 contains RCTs with the lowest positive estimated lift and decile 10 contains RCTs with the highest positive estimated lifts. For each decile we show three boxplots where the left-most boxplot summarizes the RCT lifts. Since we form deciles based on these RCT lifts the interquartile range is small. The middle and right boxplots for each decile summarize the lifts estimated by SPSM and DML, respectively. 
Because the boxplots are difficult to interpret for the lowest deciles, we also summarize the medians, 25th and 75th percentiles in Table~\ref{Lift_RCT_SPSM_DML_Box.table}. 

\begin{table}[!htbp] \centering 
\begin{threeparttable}
  \caption{Comparison of RCT Lifts with Lifts Estimated using SPSM and DML$^\dag$} 
  \lspace{1}
\label{Lift_RCT_SPSM_DML_Box.table} 
\small 
\begin{tabular}{@{\extracolsep{0pt}} ccc|ccc|ccc} 
\\[-1.8ex]\hline 
\hline \\[-1.8ex] 
 & &  RCT & \multicolumn{3}{c}{SPSM} &  \multicolumn{3}{c}{DML} \\  \cmidrule(lr){3-3} \cmidrule(lr){4-6} \cmidrule(lr){7-9} 
Funnel & RCT Decile & p50 & p25 & p50 & p75 & p25 & p50 & p75 \\ \hline
 & 1 & -0.005 & 0.099 & 0.317 & 0.699 & 0.038 & 0.232 & 0.640 \\ 
 & 2 & 0.007 & 0.104 & 0.266 & 0.466 & 0.062 & 0.220 & 0.428 \\ 
 & 3 & 0.025 & 0.266 & 0.489 & 0.954 & 0.082 & 0.390 & 0.805 \\ 
 & 4 & 0.081 & 0.529 & 0.906 & 1.589 & 0.207 & 0.609 & 1.274 \\ 
Upper & 5 & 0.188 & 0.648 & 1.375 & 2.469 & 0.390 & 0.862 & 2.051 \\ 
Funnel & 6 & 0.347 & 1.240 & 2.144 & 4.020 & 0.870 & 1.694 & 3.053 \\ 
 & 7 & 0.513 & 1.393 & 2.459 & 5.425 & 0.746 & 1.885 & 3.887 \\ 
 & 8 & 0.768 & 1.896 & 4.180 & 6.528 & 0.962 & 2.791 & 5.544 \\ 
 & 9 & 1.274 & 4.680 & 8.524 & 13.134 & -0.271 & 3.378 & 9.297 \\ 
 & 10 & 2.349 & 4.403 & 14.656 & 19.512 & -0.843 & 2.814 & 16.340 \\ \hline
 & 1 & -0.006 & 0.244 & 0.433 & 0.990 & 0.039 & 0.406 & 0.840 \\ 
 & 2 & 0.010 & 0.140 & 0.438 & 0.902 & -0.081 & 0.402 & 0.805 \\ 
 & 3 & 0.030 & 0.199 & 0.829 & 1.607 & -0.074 & 0.152 & 0.843 \\ 
 & 4 & 0.065 & 0.164 & 0.595 & 1.507 & -0.058 & 0.339 & 1.080 \\ 
Mid & 5 & 0.115 & 0.62 & 1.426 & 2.615 & 0.066 & 0.511 & 1.085 \\ 
Funnel & 6 & 0.214 & 1.698 & 3.567 & 6.564 & 0.263 & 1.162 & 3.860 \\ 
 & 7 & 0.379 & 1.236 & 2.166 & 6.382 & -0.960 & 0.471 & 1.565 \\ 
 & 8 & 0.558 & 1.580 & 5.523 & 9.715 & -1.001 & 0.232 & 1.599 \\ 
 & 9 & 0.945 & 2.942 & 5.599 & 9.751 & 0.065 & 2.427 & 4.835 \\ 
 & 10 & 2.113 & 7.143 & 13.691 & 24.588 & -0.961 & 3.052 & 6.919 \\ \hline
 & 1 & -0.014 & 0.115 & 0.327 & 0.663 & -0.181 & 0.123 & 0.503 \\ 
 & 2 & 0.002 & 0.141 & 0.384 & 0.609 & -0.041 & 0.289 & 0.536 \\ 
 & 3 & 0.010 & -0.058 & 0.249 & 1.031 & -0.418 & 0.077 & 0.619 \\ 
 & 4 & 0.021 & 0.016 & 0.302 & 1.059 & -0.411 & 0.201 & 0.820 \\ 
Lower & 5 & 0.037 & 0.010 & 0.316 & 0.969 & -0.624 & -0.069 & 0.407 \\ 
Funnel & 6 & 0.075 & 0.122 & 0.460 & 1.833 & -0.392 & 0.087 & 0.766 \\ 
 & 7 & 0.145 & 0.195 & 0.625 & 1.991 & -0.046 & 0.173 & 0.688 \\ 
 & 8 & 0.293 & 0.429 & 1.565 & 3.527 & -0.437 & 0.265 & 1.328 \\ 
 & 9 & 0.602 & 0.638 & 2.411 & 5.767 & -0.006 & 0.629 & 1.822 \\ 
 & 10 & 1.474 & 1.752 & 3.433 & 10.016 & 0.741 & 1.680 & 2.825 \\ 
\hline \hline\\[-1.8ex] 
\end{tabular} 
	\begin{tablenotes}
	 	\item [\dag] Table excludes the top 1\% lifts for each position in the purchase funnel. p25, p50, and p75 correspond to the 25th, 50th, and 75th percentiles within the corresponding RCT decile.
	\end{tablenotes}
\end{threeparttable}
\end{table} 

Scanning across upper, mid, and lower funnel results in Figure~\ref{Lift_RCT_SPSM_DML.tex} reveals a few key results. First, SPSM overestimates the RCT lift by a large amount. To see this result more easily, we characterize the performance using the absolute percentage error (APE) between a given method $m$'s lift estimate and the RCT lift estimate as
\begin{align}\label{APE}
\mbox{APE}^{m}& = \left | \frac{\ell^{m}-\ell^{rct}}{\ell^{rct}} \right | \ .
\end{align}

For example, suppose an RCT yields an RCT lift estimate of 10\% and an observational lift estimate of 50\%. We would say that the observational method overestimates by a factor of 5. Conversely, if the observational estimate was 5\%, we would say that the observational model underestimates by a factor of 2. Table~\ref{Lift_RCT_SPSM_DML_overestimation.table} summarizes the degree to which SPSM and DML over- or underestimate RCT lifts, expressed as the APE by decile of the RCT lift. Because APE is difficult to interpret for non-positive RCT lifts, the following table summarizes results only for positive RCT lifts (this excludes 10.04\% of RCTs). 

\begin{table}[!htbp] \centering 
\begin{threeparttable}
  \caption{Absolute percentage error (APE) between SPSM/DML and RCT lifts$^\dag$} 
  \lspace{1}
\label{Lift_RCT_SPSM_DML_overestimation.table} 
\small 
\begin{tabular}{crrrrrr}  
\\[-1.8ex]\hline 
\hline \\[-1.8ex] 
 & \multicolumn{2}{c}{Upper Funnel} &  \multicolumn{2}{c}{Mid-Funnel} &  \multicolumn{2}{c}{Lower Funnel} \\  \cmidrule(lr){2-3} \cmidrule(lr){4-5} \cmidrule(lr){6-7}
RCT Decile & SPSM & DML & SPSM & DML & SPSM & DML \\ \hline\hline
1 & 5393\% & 5202\% & 5727\% & 8750\% & 12905\% & 18694\% \\ 
2 & 1912\% & 2244\% & 3077\% & 2452\% & 4061\% & 4528\%\\ 
3 & 1231\% & 1252\% & 1255\% & 1139\% & 1944\% & 3222\%\\ 
4 & 687\% & 455\% & 930\% & 466\% & 928\% & 1646\% \\ 
5 & 556\% & 485\% & 954\% & 597\% & 550\% & 885\% \\ 
6 & 346\% & 339\% & 877\% & 409\% & 665\% & 503\%\\ 
7 & 468\% & 255\% & 734\% & 265\% & 435\% & 301\% \\ 
8 & 582\% & 294\% & 972\% & 248\% & 384\% & 175\%\\ 
9 & 473\% & 196\% & 406\% & 297\% & 96\% & 86\%\\ 
10 & 306\% & 168\% & 451\% & 150\% & 200\% & 77\%\\ \hline
Median & 696\% & 557\% & 948\% & 488\% & 764\% & 672\% \\  
\hline \\[-1.8ex] 
\end{tabular} 
	\begin{tablenotes}
	 	\item [\dag] Table contains results only for positive RCT Lifts.
	\end{tablenotes}
\end{threeparttable}
\end{table} 
For example, among RCTs with an upper funnel lift in the 5th decile, the table shows that the median overestimate of SPSM lift relative to RCT lift is 556\%. Looking across the table, we observe that SPSM overestimates the RCT lift by about a factor of 5,400\% to 12,900\% for RCTs in the 1st decile and by a factor of 306\% to 451\% for RCTs in the 10th decile. For the median RCT, SPSM overestimates the RCT lift by a factor of 696\%, 948\%, and 764\% for upper, mid, and lower funnel outcomes, respectively. 

The second key result is that DML overestimates the RCT lift somewhat less. For the median RCT, DML overestimates the RCT lift by a factor of 557\%, 488\%, and 672\% for upper, mid, and lower funnel outcomes, respectively. 

One problem in interpreting the summaries in Table~\ref{Lift_RCT_SPSM_DML_overestimation.table} is that they are multiplicative and thus tend to be large for RCT's in the lower deciles with smaller lifts. Hence, Table~\ref{Lift_RCT_SPSM_DML_absdiff.table} reports instead the absolute error (AE) of SPSM and DML, respectively. The absolute error of a given method, $m$, is
\begin{align}\label{AE}
\mbox{AE}^{m}& = \left | \ell^{m}-\ell^{rct} \right | \ .
\end{align}

\begin{table}[!htbp] \centering 
  \caption{Absolute error (AE) between SPSM/DML and RCT lifts} 
  \lspace{1}
\label{Lift_RCT_SPSM_DML_absdiff.table} 
\small 
\begin{tabular}{crrrrrr}
\\[-1.8ex]\hline 
\hline \\[-1.8ex] 
 & \multicolumn{2}{c}{Upper Funnel} &  \multicolumn{2}{c}{Mid-Funnel} &  \multicolumn{2}{c}{Lower Funnel} \\  \cmidrule(lr){2-3}  \cmidrule(lr){4-5}  \cmidrule(lr){6-7}
RCT Decile & SPSM & DML & SPSM & DML & SPSM & DML \\ \hline\hline
1 &30\%&40\%&45\%&59\%&37\%&35\%\\ 
2 &27\%&28\%&46\%&52\%&44\%&40\%\\ 
3 &49\%&52\%&79\%&53\%&37\%&49\%\\ 
4 &78\%&69\%&54\%&51\%&35\%&62\%\\ 
5 &135\%&94\%&152\%&71\%&38\%&56\%\\ 
6 &182\%&141\%&281\%&117\%&38\%&54\%\\ 
7 &262\%&148\%&257\%&137\%&54\%&46\%\\ 
8 &499\%&258\%&503\%&158\%&139\%&85\%\\ 
9 &677\%&296\%&378\%&299\%&175\%&79\%\\ 
10 &1441\%&866\%&1499\%&492\%&267\%&129\%\\ \hline
Median &128\%&115\%&152\%&103\%&56\%&57\%\\    
\hline \\[-1.8ex] 
\end{tabular} 
\end{table} 

For example, where ``RCT Decile'' equals 5, the value of 135\% under Upper Funnel (SPSM) indicates that the median AE between RCT lift and SPSM lift is 135 percentage points. This median is calculated over the set of RCTs in the 5th decile of all RCTs with upper funnel outcomes. This metric confirms that DML leads to less mis-estimation than SPSM for upper- and mid-funnel outcomes. SPSM and DML are comparable for lower funnel outcomes according to the AE metric. 

The third key result is that the interquartile range of DML estimates is smaller than that of SPSM estimates. This is true in particular for mid and lower funnel outcomes. 

The fourth key result is that DML, while generally superior to SPSM, still fails to reliably approximate the RCT estimates. The median absolute percentage point difference (AE) between RCT and DML lift estimates is 115\%, 103\%, and 57\% for upper, mid, and lower funnel outcomes, respectively. These are very large measurement errors, given that the median RCT lifts are 29\%, 18\%, and 5\% for the equivalent funnel outcomes, respectively.\footnote{In comparing these absolute errors with the lift estimates reported in the abstract and introduction, please note that the median of the absolute difference is not the same as the difference of the medians.} 

\subsection{Reducing Selection Bias}

The poor performance of observational methods raises the question of whether these methods are reducing the bias inherent from comparing exposed and unexposed users, i.e., whether the methods improve upon the ``exposed-unexposed'' estimate. If these methods and the large feature set being used do not lead to better performance, it may indicate that observational methods are mis-specified or that the features we are utilizing are not useful in explaining selection.

\begin{figure}
 	\centering
	\caption{Absolute Percentage Error by Event Funnel}
	\label{spsm_dml_naive_are_comparison.tex}
	\input{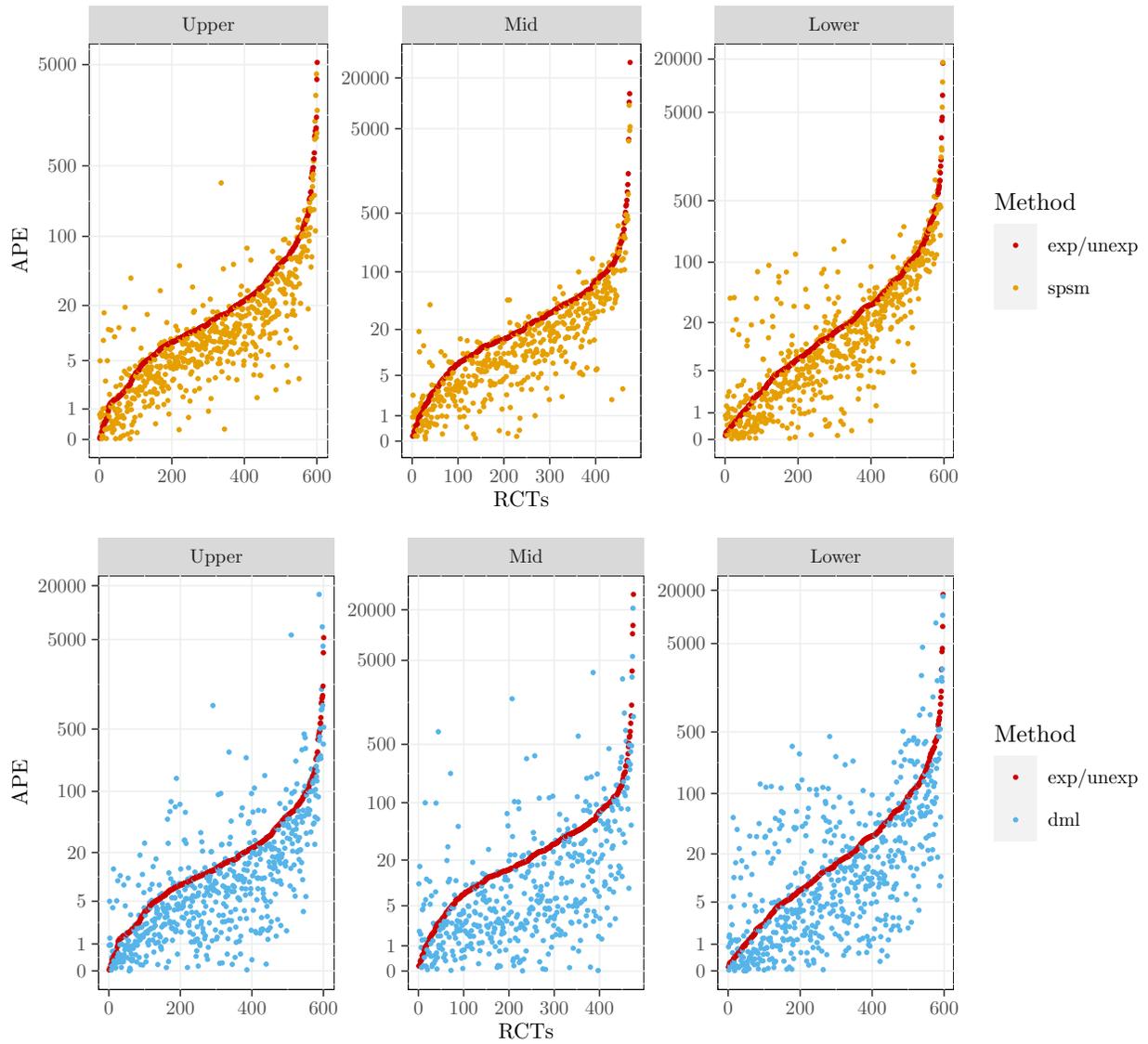}
\end{figure}

In Figure~\ref{spsm_dml_naive_are_comparison.tex} we sort all RCTs from the lowest APE when using the exposed-unexposed lift estimate to the largest APE. We then plot the APE for the exposed-unexposed and for the SPSM estimates by purchase funnel position.
The top panel in the figure shows that SPSM improves upon the exposed-unexposed estimate in the majority of cases. The APE of the SPSM estimate nearly always falls below the APE of the exposed-unexposed estimate. In other words, SPSM reduces the selection bias that arises from non-random assignment into the exposed and unexposed groups. However, it does not manage to reduce the bias enough to come close to the RCT estimates.

The bottom panel in Figure~\ref{spsm_dml_naive_are_comparison.tex} shows that DML often reduces the APE of an RCT more than SPSM does, however, compared to SPSM there are more cases where DML increases the APE compared to the exposed-unexposed estimate. 

\begin{figure}
 	\centering
	\caption{Remaining Percentage Bias by Purchase Event Position}
	\label{remaining_bias_by_funnel.tex}
	\input{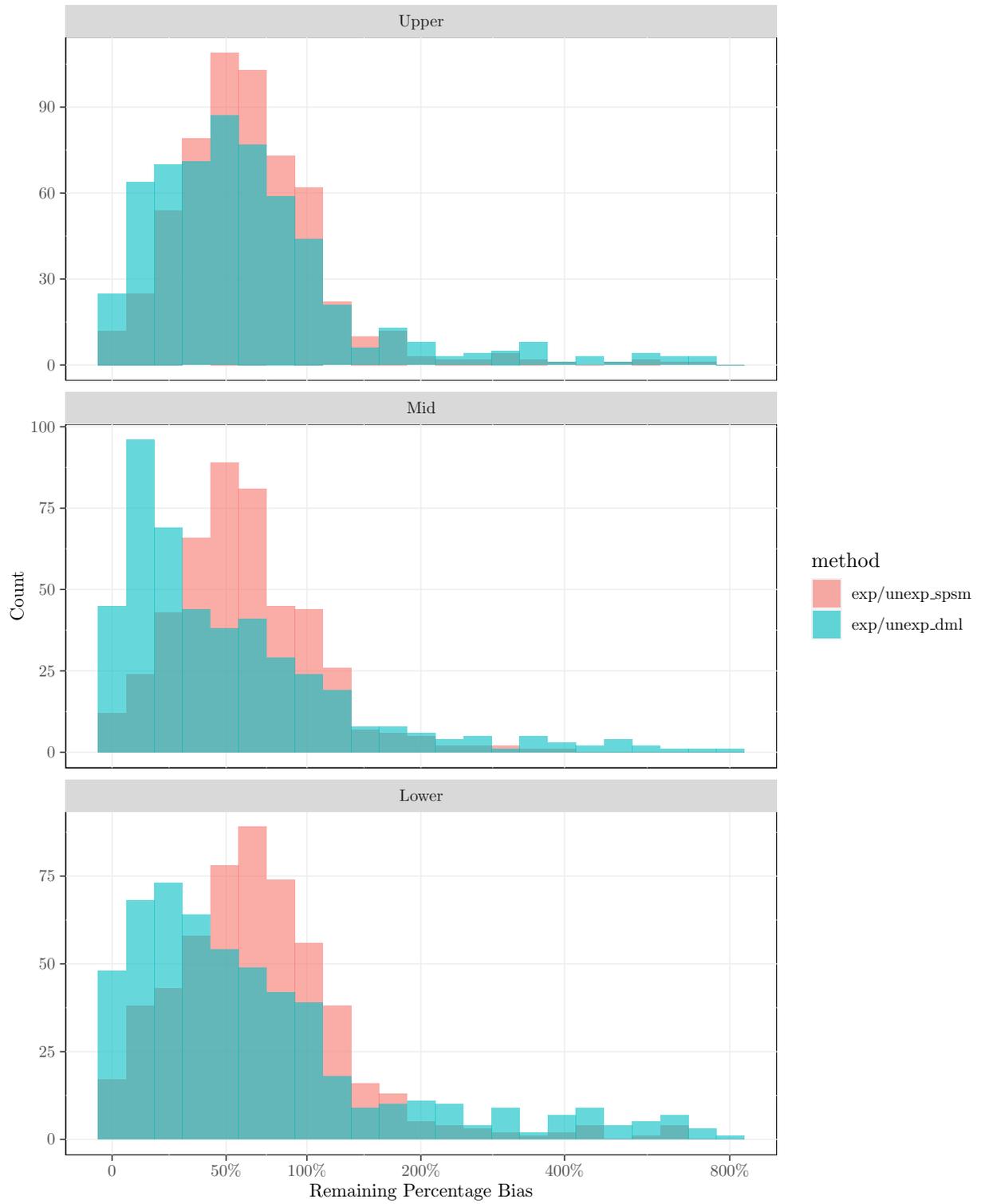}
\end{figure}

We can formalize the degree to which SPSM and DML manage to reduce the APE of the exposed-unexposed estimate by calculating the remaining percentage bias (RPB), i.e., the remaining selection effect, after each method has tried to reduce the total selection effect inherent in the exposed-unexposed ($\eu$) estimate with 
\begin{align}
\mbox{RPB}^{m} & = \left (1-\frac{\mbox{APE}^{\eu}-\mbox{APE}^{m}}{\mbox{APE}^{\eu}} \right )*100 
\end{align}
where $\mbox{m} \in \mbox{\{SPSM, DML\}}$. Figure \ref{remaining_bias_by_funnel.tex} reports the RPB estimates by funnel position. The mean, median, and mode of the remaining percentage bias distribution for all three purchase funnel positions are below 100\%. This finding shows that SPSM and DML usually reduce the selection bias present in these RCTs. However, RPB is not bound by 100\%. In some cases the predictive models underlying SPSM or DML estimates may be noisy or trying to estimate a small effect which can result in situations where exposed-unexposed estimates may actually provide a closer estimate of RCT results.

\begin{table}[!htbp] \centering 
  \caption{Percent of RCTs with RPB Below Threshold} 
  \lspace{1}
\label{remaining_relative_bias_cutoffs.table} 
\small 
\begin{tabular}{@{\extracolsep{0pt}} ccccc} 
\\[-1.8ex]\hline 
\hline \\[-1.8ex] 
Method & Funnel Position & RPB \textless  100\% & RPB \textless  50\% & RPB \textless  20\% \\ 
\hline \\[-1.8ex] 
DML & Upper & 82.8\% & 52.1\% & 20\% \\ 
DML & Mid & 84\% & 66.1\% & 42.2\% \\ 
DML & Lower & 79.3\% & 60.1\% & 30.6\% \\ 
SPSM & Upper & 87.1\% & 39.5\% & 8.4\% \\ 
SPSM & Mid & 84.9\% & 42.7\% & 10.3\% \\ 
SPSM & Lower & 80.7\% & 41.1\% & 14.7\% \\ 
\hline \\[-1.8ex] 
\end{tabular} 
\end{table}

However, as Table~\ref{remaining_relative_bias_cutoffs.table} shows, the overall performance of SPSM and DML in reducing selection bias is relatively poor: SPSM improves on the exposed-unexposed estimate in 79.7\% to 86.7\% of RCTs, depending on the purchase funnel position. Consistent with the lower panel of Figure~\ref{spsm_dml_naive_are_comparison.tex}, DML does a bit worse: it improves on the exposed-unexposed estimate in 77.9\% to 82.2\% of RCTs, depending on the purchase funnel position. However, DML does better than SPSM at reducing the selection bias by 50\% or more (51\% to 59\% vs. 39\% to 42\%) and at reducing the selection bias by 80\% or more (19\% to 42\% vs. 8\% to 13\%).

These results lead to two main findings. First, observational models such as SPSM and DML are better at estimating RCT results when compared to a simple comparisons of exposed and unexposed users. In scenarios where treatment is not randomly assigned, methods that control for self-selection into treatment led to noticeable improvements in the accuracy of causal estimates. Second, we observe that even though there is variation in performance across the types of observational models and position in the purchase funnel, observational methods still regularly produce estimates that are unable to accurately estimate causal effects. Even in the best cases observational methods were unreliable.

\section{Explaining Performance} \label{sec: Explaining Performance}

The previous subsection shows considerable heterogeneity in how well SPSM and DML can measure the causal effect of ad campaigns. While SPSM and DML lift estimates are, on average, much higher than RCT lift estimates, this is not always true. Ad campaigns are described by a wide range of features that are either predefined or chosen by an individual advertiser such as the industry vertical an advertiser is part of, the baseline conversion rate that advertisers could expect without ads, the way they target and deliver ads, and the size of a campaign. If there are segments of ad campaigns where observational models perform well, it is possible that some advertisers could utilize these approaches to obtain causal estimates on their campaign's effectiveness. To understand what these segments look like, we now investigate whether the characteristics of the ad campaign explain when SPSM and DML perform comparatively better.

To see how characteristics of the ad campaign are correlated with SPSM and DML performance, we build a predictive model with the $\mbox{APE}$ of an observational method as the target and using characteristics of the RCT as features. The unit of observation is an RCT. We include the following characteristics: The length of the experiment (in days), the number of users in the test group, the conversion rate of the control group, the exposure rate (what percent of targeted users in the test group were exposed to the ad), the position of the conversion outcome in the purchase funnel, the out-of-sample predictive performance of the propensity scoring model (measured as ``area under the ROC curve'' or ``AUC''), and the prospecting ratio (which is 1 if all targeted users were prospects, 0 if all targeted users were chosen for remarketing, and intermediate values for a mixture of prospecting and remarketing). For DML we also include the out-of-sample predictive performance of the outcome model (measured as AUC). 

We implement a 95\% winsorization by setting each $\mbox{APE}$ above the 95th percentile in each purchase funnel position to the values of that 95th percentile (for SPSM, this affects APEs above 122 for upper-funnel outcomes, 157 for mid-funnel outcomes, and 230 for lower-funnel outcomes; for DML this affects APEs above 149 for upper-funnel outcomes, 250 for mid-funnel outcomes, and 297 for lower-funnel outcomes).

To account for non-linearities and interactions, we use a random forest. Since this model allows for interactions, we train the model on all RCTs (for the two observational methods across the three purchase funnel positions). Tuning the model yields two randomly selected features at each cut in the tree and a minimum node size of one. We use permutation importance to evaluate the relative importance of each explanatory variable. The results are scaled from 100 (most important feature) to 0 (least important feature) and shown in Table~\ref{rf_variable_importance_combined_APE.table}.\footnote{We use ``permutation importance,'' as proposed by Breiman and Cutler. The process consists of establishing a baseline $R^2$ (for a continuous outcome variable) by putting the test samples down the random forest. Next, for each experiment characteristic, in turn, permute its values, put the permuted data down the random forest, and recompute $R^2$. Finally, let the importance of an experiment characteristic be the drop in $R^2$ caused by having permuted that experiment characteristic. The more $R^2$ drops after permuting, the more important the experiment characteristic. See \url{https://www.stat.berkeley.edu/~breiman/RandomForests/cc_home.htm}}

\begin{table}[!htbp] \centering 
  \caption{Variable Importance for SPSM and DML (Random Forest explaining APE)} 
  \lspace{1}
\label{rf_variable_importance_combined_APE.table} 
\footnotesize 
\begin{tabular}{@{\extracolsep{0pt}} lclc} 
\\[-1.8ex]\hline 
\hline \\[-1.8ex] 
Relative Variable Importance & SPSM & Relative Variable Importance & DML \\ 
\hline \\[-1.8ex] 
Prospecting (vs. Remarketing) Ratio & $100$ & Prospecting (vs. Remarketing) Ratio & $100$ \\ 
Control Group Conversion Rate & $88.4$ & AUC from Outcome Model & $97.4$ \\ 
Lower Funnel Dummy & $66.8$ & Exposure Rate & $92.7$ \\ 
Exposure Rate & $51.8$ & Experiment Length (in days) & $79.3$ \\ 
Number of Users in Test Group & $49.2$ & Number of Users in Test Group & $75.4$ \\ 
AUC from Propensity Score Model & $48.1$ & Control Group Conversion Rate & $72.8$ \\ 
Experiment Length (in days) & $43.5$ & AUC from Propensity Score Model & $66.6$ \\ 
Mid-Funnel Dummy & $0$ & Lower Funnel Dummy & $27.8$ \\ 
 & $$ & Mid-Funnel Dummy & $0$ \\ 
\hline \\[-1.8ex] 
\end{tabular} 
\end{table}

The most important feature for explaining absolute percentage error for both DML and SPSM is the prospecting ratio. Otherwise, the models differ in the order of variable importance. One of the most important features for SPSM, Lower Funnel Dummy, is towards the very bottom of the ranking for DML. In the feature importance ranking, the scaled feature importance scores drop significantly after the second or third feature for SPSM but tend to decrease more uniformly for DML.

\begin{figure}
 	\centering
	\caption{Partial Dependence Plots (Random Forest explaining APE)}
	\label{pdp_rf_A.tex}
	\includegraphics[height=7.5in]{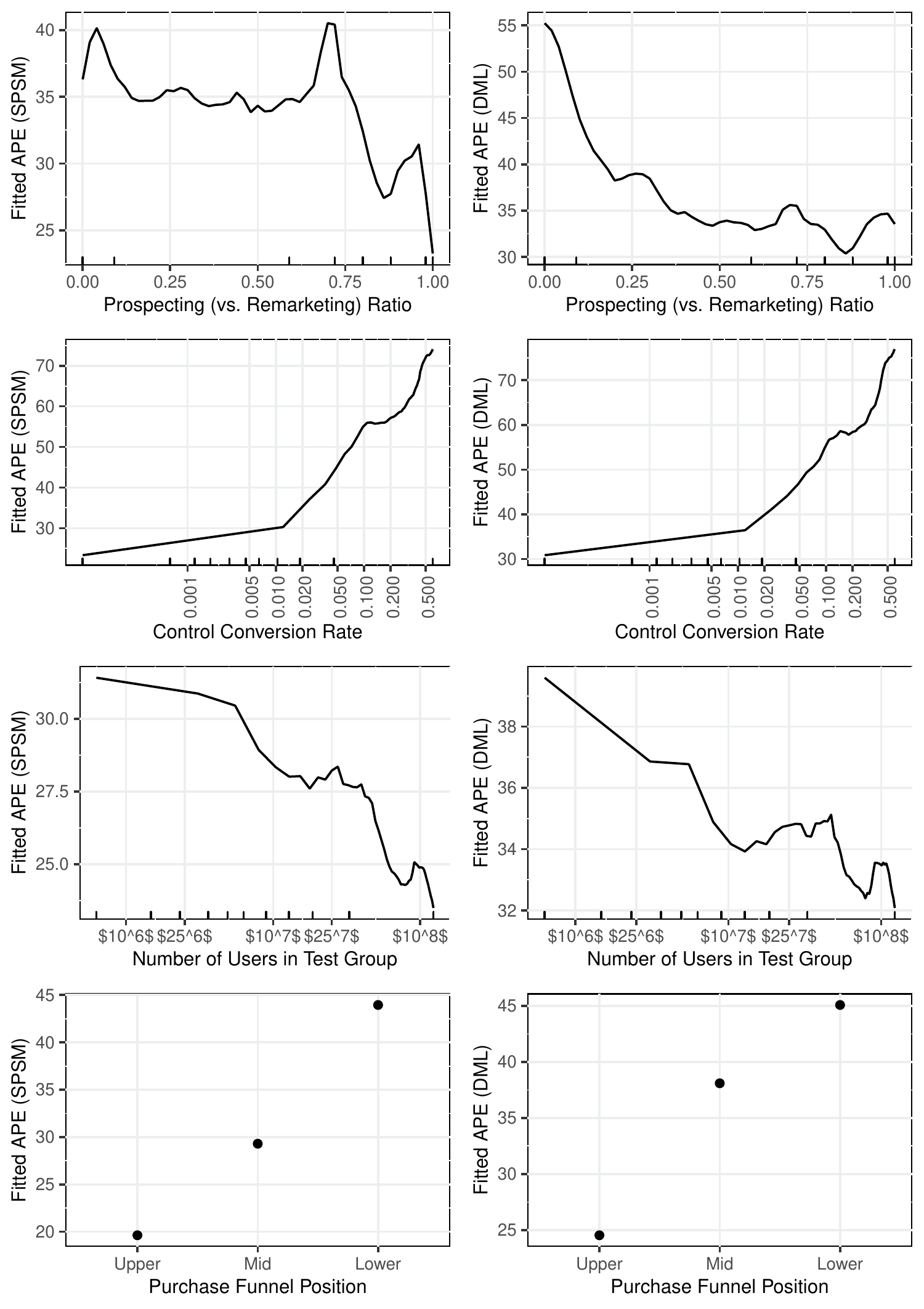}
\end{figure}

\begin{figure}
 	\centering
	\caption{Partial Dependence Plots (Random Forest explaining APE)}
	\label{pdp_rf_B.tex}
	\includegraphics[height=7.5in]{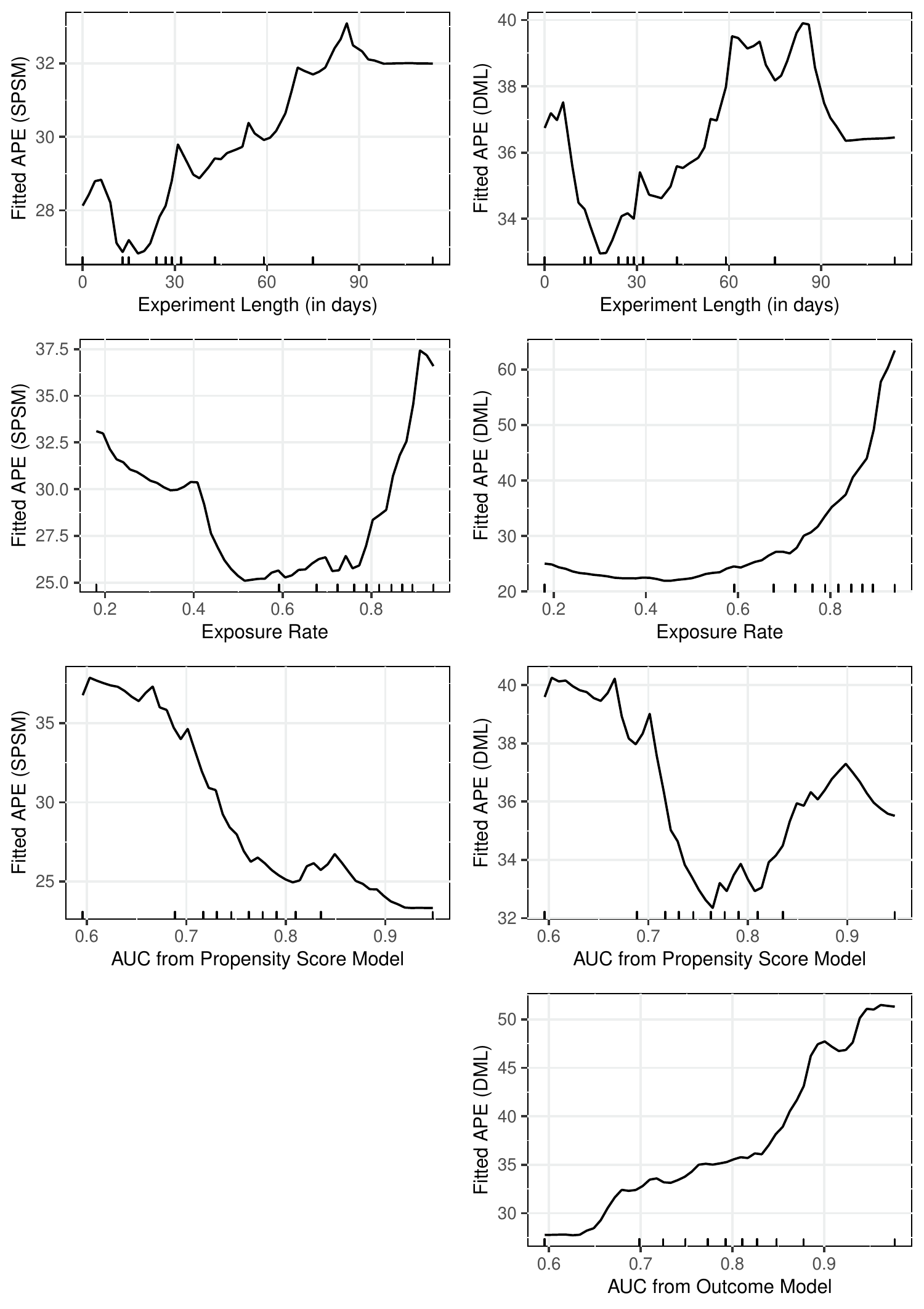}
\end{figure}

To better evaluate the relationship between experiment characteristics and the performance of SPSM and DML, Figures~\ref{pdp_rf_A.tex} and ~\ref{pdp_rf_B.tex} visualize the relationship between each experiment characteristic and $\mbox{APE}^{SPSM}$ and $\mbox{APE}^{DML}$ while accounting for the average effect of the other experiment characteristics in the model. These plots are referred to as partial dependence plots (PDP) \citep{Greenwell2017}. Note that each plot displays a ``rug'' along the x-axis, indicating deciles of the observations for each experiment characteristics. In these plots, the PDP for SPSM are graphed on the left while the plots for DML are graphed on the right. We sort the plots by decreasing variable importance of each variable in the SPSM model. 

To interpret the PDP of prospecting ratio, notice the rug: half of the RCTs targeted a majority of users by prospecting only (no remarketing). The APE of these experiments is lower than those of experiments that target most users through remarketing, for both SPSM and DML. We speculate that remarketing introduces additional endogeneity, which the observational methods have difficulty correcting. Next, we consider the conversion rate of the control group. The smaller that conversion rate, the better SPSM and DML perform. We believe the reason to be straightforward: If untreated users do not convert, no selection effect needs to be corrected by the observational methods. Third, it seems that a larger number of users in the test group is associated with a lower APE, except for the very largest number of users for DML. Perhaps a larger sample helps the selection model. Fourth, we find that both observational methods perform better for upper-funnel than for mid- or lower funnel purchase outcomes. We think this is because there is more scope for selection for lower-funnel outcomes.

Moving to Figure~\ref{pdp_rf_B.tex}, we do not find a clear relationship between experiment length and model performance. Next, we find that worse APEs are associated with higher exposure rates in the test group. We suspect this is true because a higher exposure rate implies a smaller proportion of unexposed users that can be used to predict counterfactual outcomes for the treated users. We also find that for SPSM a better-fitting propensity model (higher AUC) is associated with lower APEs. We do not have a prior about this relationship since a propensity score model should not perfectly predict. In fact, we do not see the same pattern for DML. Finally, surprisingly, we do not find a monotonic relationship between the the AUC of the outcome model and model performance. 

Overall, SPSM and DML seem to work better in some settings. However, even in these settings, the average percentage error of both observational methods remains high. 

% May 1, 2022: Here is the OLS "hedonic regression" table to summarize the PDPs, requested by R2 point 6.1
%\input{OLS_explaining_APE.table}

\section{Conclusion} \label{sec: Conclusion}
This paper makes three contributions. First, we are the first to characterize the effectiveness of Facebook advertising using a collection of experiments that are representative of the large-scale RCTs advertisers run on Facebook in the United States. The median RCT lifts are 29\%, 18\%, and 5\% for upper, mid, and lower funnel outcomes, respectively. We found that 76\%, 74\%, and 60\% of these RCT lifts are statistically different from zero across their respective funnel outcomes. The fraction of insignificant results can be partially explained by the fact that 25\% of RCTs were not sufficiently powered to detect a lift of 10\%. Moreover, we described ad effectiveness by industry vertical. These results could serve as useful prior distributions to aid firms' digital advertising decisions, complementing the results on TV ad effects in \cite{ShapiroHitschTuchman2021}.

Second, we add to the literature on whether non-experimental data are ``good enough'' to power  advertising measurement. Specifically, we used data from 1,673 campaigns at Facebook, each of which was described using over 5,000 user- and experiment-level features, to implement stratified propensity score matching and double/debiased machine learning. We find that, on average, SPSM and DML both overestimate the RCT lift by a large amount. While DML performs better than SPSM, a more traditional program evaluation approach, it still fails to reliably approximate the RCT estimates. The median absolute percentage point difference (AE) between RCT and DML lift estimates is 115\%, 103\%, and 57\% for upper, mid, and lower funnel outcomes, respectively. These are very large measurement errors, given that the median RCT lifts are 29\%, 18\%, and 5\% for the respective funnel outcomes. 

Third, we characterize the circumstances under which SPSM and DML perform better or worse at recovering the causal effect of advertising. While both approaches seem to work better in some settings, even in those cases, the average percentage error obtained with either model, using the data at our disposal, remains high.

The models and data we use surpass what individual advertisers are able to use for ad measurement and represent close to the peak of what third-party measurement partners and large advertising platforms currently employ. Nonetheless, despite the quality of the data available and the flexibility of the models employed, we found these were inadequate to consistently control for the selection effects induced by the advertising platform. 

\subsection{Next Steps for Researchers and Industry}

Experimentation platforms are appealing because they offer robust and elegant solutions for ad measurement. However, we think there are a number of possible paths forward for improving advertising measurement.

The first path requires advertising platforms to alter their data logging and retention practices to meet the requirements of observational methods. This path requires recording and operationalizing vastly more data than is currently done in practice, such as all the features that generate ad rankings at the user-bid-request level. In addition, this path would require a selection model at the user-bid-request, as opposed to only at the user level. This is similar in concept to the approach that \cite{tunuguntla_2021} pursues, using a custom Demand Side Platform to observe bid-level data and a user's session history. More evidence is needed to understand the efficacy of this approach on more sophisticated ad platforms and across more advertising campaigns. As we note in the Introduction, such a logging system would require significant engineering and storage investments to operate at this scale---it may be difficult for companies to justify these investments for measurement. Another issue is that ad delivery systems are actually an ensemble of systems, typically managed by different teams. Solutions that rely on this collection of systems could be brittle in the sense that updates to one part of the system might break the overall method.

The second path is to identify pre-existing sources of quasi-randomness in the ad delivery process. Although some additional data logging might be required, these solutions would likely not require significant modifications to existing delivery processes. One example of this is the probabilistic inclusion of advertisers in an auction due to budget pacing rules \citep{gui_nair_niu_2022}. The downside of this approach is that the effects obtained may not always correspond to those of interest to the advertiser or the platform.

A third path is to inject exogenous variation in advertising exposure \emph{for the purpose of building scalable models for ad measurement.} This randomness could be added at different levels. For example, one possibility is to implement RCTs for a subset of campaigns, and then to model the relationship between the RCT lift and a set of easily observed non-causal summary, or ``proxy'', metrics. \cite{lift3} pursue this approach, reframing the causal inference problem as a prediction problem where the unit of observation is the campaign. They show that attribution-style metrics, such as last-click conversions, can be predictive of incremental lift. Exogenous variation can also be introduced at more granular levels, such as the bid-level randomization implemented in \cite{lewis_wong_2018} and  \cite{waisman2022}. These two methods have the benefit of estimating the incremental effects of impressions while optimizing the advertiser's real-time bidding strategy.

A fourth path, which is complementary to all of the above, is to devote more effort to understand how advertisers make decisions based on their advertising measurements. In this paper, we focused on how close the estimates from observational data came to those from RCTs. However, it is possible that an advertiser would make the same decision regarding a campaign with either of these estimates. That is, a biased causal effect does not necessarily lead a decision maker to the wrong decision \citep{loria_provost_2022}. Two challenges are that there is no one-size-fits-all model of advertiser decision making and incentive conflicts plague the industry \citep{gordon_et_al_2020_jm}. More work is needed to integrate causal effects estimation in actionable models of advertising decisions.

%%%%%%%%%%%%%%%%%%%%%%%%%%%%%%%%%%%%%%%%%%%%%%%%%%
% END OF BODY
%%%%%%%%%%%%%%%%%%%%%%%%%%%%%%%%%%%%%%%%%%%%%%%%%%

\newpage

\section*{Funding and Competing Interests}

To be allowed to access the data required for this paper, Gordon and Zettelmeyer were part-time employees of Facebook with the title of Academic Researchers, employed for three hours per week. Moakler is an employee of Meta Platforms, Inc. and owns stock in the company. 

\newpage
\lspace{1.1}
%\defaultbibliography{refs}
\bibliographystyle{chicago}
\bibliography{fbwhite}

\clearpage

\lspace{1.1}
\setcounter{equation}{0}
\setcounter{page}{1}
\setcounter{table}{0}
\setcounter{figure}{0}
\setcounter{section}{0}
\renewcommand{\thepage}{A-\arabic{page}}
\renewcommand{\thetable}{A-\arabic{table}}
\renewcommand{\thefigure}{A-\arabic{figure}}
\renewcommand{\thesubsection}{\arabic{subsection}}

\section*{Appendix}\label{Appendix}

\begin{table}[!htbp] \centering 
  \caption{Deciles of ATTs by Funnel Level} 
  \lspace{1}
\label{deciles_att_event_funnel.table} 
\small 
\begin{tabular}{@{\extracolsep{0pt}} cccc} 
\\[-1.8ex]\hline 
\hline \\[-1.8ex] 
 & Upper & Mid & Lower \\ 
\hline \\[-1.8ex] 
10\% & $0.00004$ & $0.00002$ & $$-$0.00003$ \\ 
20\% & $0.0002$ & $0.0001$ & $0.00002$ \\ 
30\% & $0.001$ & $0.0002$ & $0.0001$ \\ 
40\% & $0.001$ & $0.0004$ & $0.0001$ \\ 
50\% & $0.003$ & $0.001$ & $0.0002$ \\ 
60\% & $0.005$ & $0.001$ & $0.0004$ \\ 
70\% & $0.008$ & $0.002$ & $0.001$ \\ 
80\% & $0.012$ & $0.003$ & $0.001$ \\ 
90\% & $0.023$ & $0.005$ & $0.002$ \\ 
100\% & $0.129$ & $0.022$ & $0.025$ \\ 
\hline \\[-1.8ex] 
\end{tabular} 
\end{table}

%\begin{figure}
%\centering
%\caption{Number of features required to achieve 80\% cumulative importance}
%\label{fig:feat_importance}
%\includegraphics[width=4in]{FeatImportance_80.pdf}
%\end{figure}

%\begin{figure}
%\centering
%\caption{Number of features required to achieve 80\% cumulative importance}
%\label{fig:feat_importance}
%\includegraphics[width=4in]{FeatImportance_80.pdf}
%\end{figure}
%
%
%\begin{figure}[h]
%  	\centering
%	\caption{Distribution of SPSM Over/Underestimation}
%	\label{over_under_estimate_histogram_by_vertical.tex}
%	\input{over_under_estimate_histogram_by_vertical.tex}
%\end{figure}

\end{document}